\documentclass{aa}

\usepackage{graphicx}
\usepackage{txfonts}
\usepackage{natbib,twoopt}
\usepackage{hyperref} 
\bibpunct{(}{)}{;}{a}{}{,}  

\usepackage{graphicx}	
\usepackage{amsmath,bm}	
\usepackage{orcidlink}

\begin{document}

\title{A multi-ion non-equilibrium solver for ionised astrophysical plasmas with arbitrary elemental abundances}
\titlerunning{NEQ multi-ion solver for radiation-MHD}

\author{
  Arun Mathew
  \orcidlink{0000-0001-9896-4243}\inst{1}
  \and Jonathan Mackey \orcidlink{0000-0002-5449-6131}\inst{1}
  \and Maggie Celeste \orcidlink{0009-0002-8687-2282}\inst{2}
  \and Thomas J.~Haworth \orcidlink{0000-0002-9593-7618}\inst{3}
  \and Garrelt Mellema \orcidlink{0000-0002-2512-6748}\inst{4}
}

\institute{
  Dublin Institute for Advanced Studies, DIAS Dunsink Observatory, Dunsink Lane, Dublin 15, Ireland \\ \email{arun@cp.dias.ie}
  \and
  Institute of Astronomy, University of Cambridge, Madingley Rd, Cambridge CB3 0HA
  \and
  Astronomy Unit, School of Physics and Astronomy, Queen Mary University of London, London E1 4NS, UK
  \and
  Department of Astronomy and Oskar Klein Centre, AlbaNova, Stockholm University, SE-10691 Stockholm, Sweden
}

\date{Received -; accepted -}

\abstract
  {While many astrophysical plasmas can be modelled successfully assuming ionisation and thermal equilibrium, in some cases this is not appropriate and a non-equilibrium approach is required.
  In nebulae around evolved stars the local elemental abundances may also strongly vary in space and time.}
  {Here we present a non-equilibrium multi-ion module developed for the fluid-dynamics code \textsc{pion}, describing the physical processes included and demonstrating its capabilities with some test calculations.}
  {A non-equilbrium ionisation solver is developed that allows arbitrary elemental abundances for neutral and ionised (but not molecular) gas, for the elements H, He, C, N, O, Ne, Si, S and Fe.
  Collisional ionisation and recombination, photoionisation and charge-exchange reactions are included, and ion-by-ion non-equilibrium radiative cooling is calculated based on the instantaneous ion fractions of each element.
  Element and ion mass-fractions are advected using passive scalars, operator-split from the microphysical processes.}
  {The module is validated by comparing with equilibrium and non-equilibrium calculations in the literature.
  Effects of charge exchange on ion abundances in cooling plasmas are discussed.
  Application to modelling shocks and photoionised H~\textsc{ii}\ regions is demonstrated.
  The time-dependent expansion of a Wolf-Rayet nebula is studied, including photoionisation and collisional processes, and spectral-line luminosities calculated for non-equilibrium and equilibrium plasma states.}
  {The multi-ion module enables simulation of ionised plasmas with spatially varying elemental abundances using self-consistent ion abundances and thermal evolution.
  This allows prediction of spectral lines in UV, optical, IR and X-ray even in cases where the plasma is out of ionisation equilibrium.
  }


\maketitle

\section{Introduction}\label{sec:intro}

Hydrodynamic (HD) and magnetohydrodynamic (MHD) simulation has proven to be an essential tool in the understanding of circumstellar nebulae around stars, from pioneering early work on expansion of H~\textsc{ii} regions \citep{Las66a} to models of Wolf-Rayet nebulae \citep{GarLanMac96}, colliding-wind binaries \citep{ParPitCor11} and novae \citep{BooMohPod16}.
The thermal state of the gas can be crudely approximated by either an adiabatic or isothermal equation of state in the simplest cases, but often a more detailed approach is required.
The cooling timescale, $\tau_\mathrm{c} = E / \dot{E}$ (where $E$ is the thermal energy density), by radiative processes is typically driven by electron-ion collisions for which $\dot{E}$ depends on the density, $\rho$, according to $\dot{E}\propto \rho^2$, resulting in  $\tau_\mathrm{c} \propto \rho^{-1}$.
Dense regions of plasma may have short $\tau_\mathrm{c}$ and low-density regions long  $\tau_\mathrm{c}$, meaning that within a single simulation one may have adiabatic regions, isothermal regions and intermediate regimes.
Most research in the literature addresses this problem using tabulated cooling curves that assume collisional ionisation equilibrium \citep[CIE, e.g.,][]{Sutherland_1993} or photoionisation equilbrium \citep[PIE,][]{Wiersma_2009}, but in some environments this may be a poor assumption.

In particular for molecular clouds, the formation timescales of molecules may be much longer than the dynamical timescale of the gas, and so non-equilibrium chemistry models have been developed with varying degrees of complexity \citep[e.g.][]{NelLan97, GloMac07a, WalGirNaa15, GonOstWol17}.
The aims of these models are twofold: the chemical state of the gas partially determines its radiative heating and cooling rate, and so a correctly calculated thermal state (with associated hydrodynamical effects from pressure gradients) depends on the non-equilibrium chemistry.
Furthermore, predictions for observables such as molecular line-emission depend crucially on how accurately the molecule abundances are calculated.

In ionised plasmas, the ionisation and recombination processes are generally much faster than molecule formation, and so CIE or PIE is often, but not always, a reasonable assumption.
\citet{Raga1997} developed a limited non-equilibrium ionisation-cooling model to study bow shocks, showing that multi-dimensional simulations were feasible with this approach, at least without including radiative transfer.
\citet{TesMigMas08} introduced a comprehensive non-equilibrium cooling model to \textsc{pluto} \citep{MigBodMas07}, showing the importance of non-equilibrium effects in the radiative shocks of protostellar jets, especially for predicting spectral line intensities of commonly observed ions.
\citet{DeABre12} showed that, for simulations of the multiphase interstellar medium (ISM), the cooling timescales are often shorter than recombination timescales and so a significant fraction of the ISM volume may be quite far from CIE conditions.
This has consequences for radiative cooling rates, and hence for the gas dynamics and plasma line-emission, that cannot be captured by a tabulated equilibrium cooling curve.
In situations where we may expect the gas to be far from CIE conditions, one cannot make meaningful predictions without explicitly tracking the ionisation state of the observational tracer that we are interested in modelling.

A further complication in nebulae around evolved stars is that the elemental abundances in the stellar wind may change significantly on the dynamical timescale of the nebula, and for binary systems the winds of the two stars may have dramatically different abundances \citep[e.g.][]{PolCorSte05}.
This makes a tabulated cooling prescription similar to that of \citet{TesMigMas08} impossible to implement because the elemental abundances themselves are now functions of positions and time. \citet{Eatson_2022} calculated tabulated CIE cooling curves appropriate for the surface elemental abundances of the two stars in the binary system WR\,140, and used a passive scalar to distinguish the two winds and apply the appropriate cooling rates.
This method was applied to study WR\,140 in \citet{EatPitVan22b} and \citet{MacJonBro23} to model the transition from adiabatic to radiative shocks as the stars approach each other around periastron.
This rather simplistic approach could be improved because it is not clear that the cooling plasma is in ionisation equilibrium, and photoionisation changes the ionisation balance with respect to CIE at intermediate temperatures.

\citet{FraMel94} developed a method for modelling the gas dynamics of photoionised and collisionally ionised plasmas, applied to planetary nebulae. 
This non-equilibrium method allowed for specifying elemental abundances and was used by \citet{Mellema_1995} to study the expansion of Planetary Nebulae, including comparison of synthetic spectral line emission from simulations with observed nebulae.
\citet{MelLun02} developed this code further to investigate the 1D (spherically symmetric) expansion of circumstellar nebulae for plasmas with very different elemental abundances. They solved the non-equilibrium abundances of a large set of ions, with associated radiative cooling on an ion-by-ion basis, and studied the effect this has on the shock properties and hydrodynamical evolution of the nebulae.
For Wolf-Rayet (WR) nebulae, they found a significant effect of the elemental abundances on gas dynamics.
The C$^2$-ray method \citep{Mellema_2006} introduced a method (building on work by \citealt{AbeNorMad99}) for accurately modelling photoionisation with simulations where cells may have large optical depths.
A modified version of this method was used by \citet{Mackey_2012} with an explicit integration scheme for the radiation-MHD (R-MHD) code \textsc{Pion}.
C$^2$-ray was extended by \citet{FriMelIli12} to hard radiation fields capable of singly and doubly ionising helium, but only applied to metal-poor/free gas.

Codes that model HD or MHD generally have relatively simple photoionisation schemes because of the large computational cost of multi-frequency radiative transfer and large ionisation networks.
By contrast, R-MHD codes that employ Monte-Carlo radiative-transfer such as \textsc{Torus} \citep{HarHawAcr19} and \textsc{cmacionize} \citep{2018A&C....23...40V} have detailed radiation transport and photoionisation schemes, and may be run in equilibrium or non-equilibrium modes.
Although Monte-Carlo radiative transfer scales very efficiently \citep{HarHawAcr19} these capabilities come at significant computational cost because large numbers of photon packets are required at each timestep to give accurate results.

An important consideration for a multi-ion module including photoionisation and spatially varying elemental abundances, is that it is far from simple to construct a general equilibrium solution (neither CIE nor PIE) because the number of variables that determine the ionisation state may be quite large: density, temperature, elemental abundances, radiation flux in each frequency bin and optical depth of the grid zone being integrated.
In this case it may be simpler (and potentially faster) to compute a time-dependent solution.
There exist some popular chemistry solvers in the literature suitable for modelling ionised plasmas, for example \textsc{Krome} \citep{GraBovSch14} and \textsc{Grackle} \citep{SmiBryGlo17} but there are practical and design drawbacks to using these in \textsc{Pion}.
\textsc{Grackle} is aimed mainly at the high-redshift Universe and its photoionisation capabilities do not extend to multi-ion non-equilibrium ionisation of metals.
\textsc{Krome} could in principle be used because it has methods for multi-frequency radiation fields and photoionisation of metals, but it is written in Fortran without a C++ interface.
Given that we already had the \textsc{Sundials} ODE solver \citep{hindmarsh2005sundials, gardner2022sundials} integrated into \textsc{Pion} for simpler microphysics modules, we decided to extend this and write our own multi-ion module to calculate and solve the rate equations.

In this paper our aim is to develop and test a non-equilibrium multi-ion module for ionised plasmas in the radiation-MHD code \textsc{Pion} \citep{Mackey2021}, including collisional and photoionisaton processes, and allowing spatial and temporal variations in elemental abundances.
The module can potentially be used in many contexts from large scale simulations of galaxy formation down to au-scale simulations of colliding-wind binary systems.
The main applications we aim for are binary systems, Wolf-Rayet nebulae and supernova remnants.

In Section \ref{section:methods}, we review the previous microphysics module, followed by a detailed description of the new multi-ion model, emphasizing its advanced features and improvements.
Section \ref{section:results} presents several one-dimensional MHD shock models with different flow properties, resolution, and physical processes.
The accuracy of our multi-ion photoionisation scheme is also verified by simulating HII regions and integrating to PIE in subsection \ref{subsection:hiiregion}, comparing results with literature benchmarks.
An application of the multi-ion module is presented in Section \ref{section:wind-wind}, where we model the interaction between fast and slow winds, each having distinct chemical compositions and incorporating all the physical processes discussed in this work.
In Section \ref{section:discussion}, we discuss the computational efficiency and performance of the multi-ion module in relation to the number of chemical species and OpenMP threads, and compare equililibrium vs.~non-equilibrium solutions for wind-wind interaction, and in Section \ref{section:conclusion}, we present our conclusions.

\section{Methods}\label{section:methods}

The multi-ion module is developed for \textsc{pion}, an HD and MHD grid-based simulation code with static mesh-refinement, that incorporates radiative transfer of ionising and non-ionising photons for R-HD \citep{MacLim10} and R-MHD \citep{MacLim11}.
The first public release and description of numerical methods in \textsc{pion} is described in \citet{Mackey2021}.
The pre-existing microphysics routines in \textsc{pion} assume one of the following:
\begin{enumerate}
\item CIE with Solar abundances scaled by a metallicity, and with optional heating terms for UV heating in photoionised gas.
\item CIE for a gas composed of two different elemental abundances (distinguished by a passive tracer) with different tabulated cooling rates \citep{MacJonBro23}.
\item Non-equilibrium ionisation calculation of hydrogen based on collisional ionisation and photoionisation, and radiative cooling appropriate for atomic and partially ionised gas in the diffuse ISM \citep{MacLanGva13}, denoted \texttt{MPv3}.
\item As above, but with low-temperature cooling appropriate for denser gas \citep{MacGvaMoh15} (see also \citealt{Henney_2009}), denoted \texttt{MPv5}.
\item Two-temperature isothermal model, where neutral gas is isothermal with a low temperature and photoionised gas also isothermal with a higher temperature.
This model was used for the \textsc{StarBench} tests in \citet{BisHawWil15}, denoted \texttt{MPv7}.
\end{enumerate}

These microphysics modules have been used successfully to model both photoionised and collisionally ionised plasmas around massive stars, but have a strong limitation that the elemental abundances are assumed to be uniform throughout the domain and unchanging in time.
Furthermore the ionisation states of important diagnostic ions such as O$^+$, O$^{2+}$ and others, cannot be calculated without detailed postprocessing, assuming either CIE or PIE.
This limits the predictive power of simulations using these modules, especially for making synthetic observations, and motivates us to develop the multi-ion non-equilibrium module.

In subsection \ref{section:new_chemistry} we describe the physical processes and equations solved by the new microphysics module, dubbed \texttt{NEMO}.
Subsection \ref{subsection:ion-rec} lists the sources for collisional ionisation and recombination rates, checking that CIE abundances agree with literature results.
In subsection \ref{subsection:cie_cooling} we derive the CIE cooling functions, comparing our results with established findings to confirm the accuracy of our approach.
Subsections \ref{subsection:photoionisation} and \ref{subsection:charge-exchange} provide detailed descriptions of the photoionisation and charge-exchange schemes.

\subsection{New Non-equilibrium microphysics module, \texttt{NEMO}}\label{section:new_chemistry}

The features and the scope of the newly added non-equilibrium ionisation network and cooling function to the \textsc{pion} code are described below.
We begin with the ideal inviscid MHD equations:
\begin{eqnarray}
 \frac{\partial\rho}{\partial t} + \nabla\cdot\left(\rho\vec{v}\right) & = & 0  \,
 \label{eq:CL1} \\ 
 \frac{\partial(\rho\vec{v})}{\partial {t}} + \nabla\cdot\left(\rho\vec{v}\vec{v}^T - \vec{B}\vec{B}^T \right) + \nabla\left( p + \frac{|\vec{B}|^2}{2} \right) &=& 0\,
  \label{eq:CL2} \\
 \frac{\partial\vec{B}}{\partial t} - \nabla\times\left(\vec{v}\times\vec{B}\right) &=& 0 \,
 \label{eq:CL3} \\ 
 \frac{\partial E}{\partial t} + \nabla\cdot\left[\left(E + p + \frac{|\vec{B}|^2}{2} \right)\vec{v} - \left(\vec{v}\cdot\vec{B}\right)\vec{B}\right]
 & = & S_E,   \label{eq:CL4} 
\end{eqnarray}
where the source term $S_E$ include radiative cooling and photoheating.

The ion abundances of H, He, C, N, O, Ne, Si, S, and Fe are included using passive scalars, $X_{\kappa,i}$, that obey the usual conservation equation:
\begin{equation}\label{eq:species}
 \frac{\partial\rho X_{\kappa,i}}{\partial t} + \nabla\cdot\left(\rho X_{\kappa,i}\vec{v}\right) = S_{\kappa,i} \,,
\end{equation}
where the scalar quantity $X_{\kappa,i}$ is the mass fraction of the $i$-th ion of the element $\kappa$. The ion abundances for each element satisfy $\sum_{i=0}^{Z-1} X_{\kappa,i} = X_\kappa - X_{\kappa,Z}$, where $Z$ is the atomic number of element $\kappa$, and the mass fraction of the bare ion $X_{\kappa,Z}$ is determined by a conservation equation. For elemental abundances, $X_\kappa$, the initial conditions are constrained such that $\sum_\kappa X_\kappa = 1$ everywhere. In terms of ion abundances, this gives $\sum_\kappa \sum_{i=0}^{Z} X_{\kappa,i} = 1$.

The source term  $S_{\kappa,i}$ represents ionisation and recombination into and out of ionisation state $i$, and this is evaluated by operator splitting.
The homogeneous part of the equations (advection) are solved together with the hydrodynamics, and the source term is solved in the new microphysics module, by time-integrating the rate equations for each ion together with the thermal energy density in each cell.
This coupled system of ODEs is solved using the \textsc{cvode} solver \citep{Cohen_1996} from the \textsc{Sundials} library \citep{hindmarsh2005sundials, gardner2022sundials}.
The consistent multi-species advection (CMA) scheme of \citet{PleMul99} is used to ensure element and ion mass fractions remain consistent through the avection step (since each passive tracer is advected independently of the others).
Molecules are not considered and so the modified CMA method of \citet{GloFedMac10} is not required.

The processes of ionisation and recombination include:
\begin{enumerate}
    \item Electron-impact (collisional) ionisation
    \item Radiative recombination
    \item Dielectronic recombination
    \item Charge exchange recombination
    \item Charge exchange ionisation
    \item Cosmic-ray ionisation
\end{enumerate}
Rather than integrating \(X_{\kappa,i}\), we change variables to \(Y_{\kappa,i} = X_{\kappa,i}/X_{\kappa}\), i.e., the fractional ionisation of element \(\kappa\) in ionisation state \(i\). This gives the source terms $\tilde{S}_{\kappa,i}$ ($= S_{\kappa,i} / (\rho X_{\kappa}) $) as:
\begin{multline}\label{eq:ion_bal}
    \tilde{S}_{\kappa,i} =  Y_{\kappa,i+1} n_e \alpha_{\kappa,i+1} - Y_{\kappa,i} \left( n_e \zeta_{\kappa,i}
                + n_e \alpha_{\kappa,i} + \Gamma_{\kappa,i} + \beta \delta_{i0} \right)
   \\ + Y_{\kappa,i-1} \left( n_e \zeta_{\kappa,i-1} + \Gamma_{\kappa,i-1} \right)  \,
\end{multline}
with $\zeta_{\kappa,i}(T)$ denoting the collisional ionisation rate coefficient for the $i$-th ion of the element $\kappa$, $\alpha_{\kappa,i}(T)$ representing the corresponding recombination rate, and $\Gamma_{\kappa,i}(r)$ the photoionisation rate.
The cosmic-ray ionisation rate, denoted by $\beta$, is assumed to remain constant at $1.0 \times 10^{-17}$ s$^{-1}$ for neutral species \cite{Goldsmith_1978}.
The ollisional ionisation process encompasses electron-impact ionisation and ionisation charge exchange reactions, resulting in total ionisation rate given by:
\begin{equation}
    \zeta_{\kappa,i} = \zeta_{\kappa,i}^{\rm coll} + \frac{1}{n_e}\sum_{s,q} n_{s,q} \, k_{\kappa, i}^{s,q}
\end{equation}
Similarly, recombination processes involve radiative and di-electronic processes ($\alpha_{\kappa,i}^{\rm rec}$), along with recombination charge exchange reactions, yielding a total recombination rate,
\begin{equation}
    \alpha_{\kappa,i} = \alpha_{\kappa,i}^{\rm rec} + \frac{1}{n_e}\sum_{s,q} n_{s,q} \, k_{\kappa, i}^{s,q}
\end{equation}
Here, $k_{\kappa, i}^{s,q}$ denotes the reaction rate coefficient for the reaction involving reactants $A_{\kappa}^i$ and $B_{s}^q$, where $s$ refers to the element and $q$ to the ionisation state (similar to $\kappa$ and $i$). The summation is carried out for all species that have a charge-exchange reaction with species $(\kappa,i)$.

The source step is solved in the microphysics module with equations:
\begin{equation}
\frac{d E_{\rm int}}{d t} = S_E = -(L_{\rm coll} +L_{\rm rec} + L_{\rm cool}) + Q_{\rm cr} + Q_{\rm cx} + Q_{\rm photo} \label{eq:mp1}
\end{equation}
and
\begin{equation}\label{eq:ydot}
    \frac{dY_{\kappa,i}}{d t}  = \tilde{S}_{\kappa,i}   \,,
\end{equation}
where $E_{\rm int} = p/\gamma-1$, with $\gamma$ the adiabatic index; $\gamma=5/3$ for ionised non-relativistic plasmas.

The radiative losses given by 
\begin{equation}
L_{\rm cool} = n_e\sum_{\kappa,i} n_{\kappa, i} \, \Lambda_{\kappa, i}, 
\end{equation}
involve individual radiative cooling rates $\Lambda_{\kappa, i}(n_e, T)$ per electron, per ion of the species $(\kappa,i)$, calculated using \textsc{CHIANTIPy} version 15.0 \citep{Zanna_2021, Young_2003}.
The rate includes bremsstrahlung, line radiation and two-photon radiation processes valid for a large range of temperatures, $10-10^9$ K, and electron density, $1.0 - 10^6$ cm$^{-3}$.

Additionally, we account for radiative losses due to recombination using the following expression: 
\begin{equation}
L_{\rm rec} =  0.9 k_{\rm B} n_e T \sum_{\kappa}\sum_{i=1}^{N_{\kappa}} \alpha_{\kappa,i}^{\rm rec} n_{\kappa, i}.
\end{equation}
This equation quantifies the energy loss during the recombination process, and the factor of 0.9 is an approximate fit to the value for H recombination at temperatures around $10^4$\,K \citep{Hummer_1994},
which applies because the lower-energy electrons in the thermal distribution have a higher probability of recombining.

Collisional ionisation by electrons removes energy from the gas by liberating one or more electrons from the target atom/ion through collisions. The energy lost is determined by the product of the collisional ionisation rate, denoted as $\zeta_{\kappa,i}$, and the ionisation potential $I_{\kappa, i}$ of the ion being ionised. This process is summed over all ions and can be represented as:
\begin{equation}
L_{\rm coll} = n_e \sum_{\kappa} \sum_{i=0}^{N_{\kappa}-1} I_{\kappa, i} \zeta_{\kappa,i}^{\rm coll} n_{\kappa, i}
\end{equation}

Heating consists of cosmic-ray heating, charge-exchange heating, and photoheating.
Cosmic-ray heating is represented by the equation:
\begin{equation}
Q_{\rm cr} = \langle E_{\rm cr} \rangle \, \beta \sum_{\kappa} n_{\kappa, 0}.
\end{equation}

This is based on the simple assumption of constant deposition of average ionisation energy, with $\langle E_{\rm cr} \rangle =$ 20 eV per ionisation of neutral species into the gas \citep{Goldsmith_1978}.
More significant heating arises from photoionisation, denoted as $Q_{\text{photo}}$.
In each instance of photoionisation involving the species $(\kappa, i)$ by a photon with energy $E$, the excess energy, denoted as $(E - E_{\rm th\, (\kappa, i)})$, is transferred to the liberated electron ($E_{\rm th\, (\kappa, i)}$ is the ionisation potential of species $(\kappa, i)$).
A more comprehensive explanation of the photoionisation method and its implementation can be found in Section \ref{subsection:photoionisation}.
Lastly, $Q_{\text{cx}}$ denotes the heating resulting from charge-exchange reactions.
Charge exchange reactions emerge as the primary contributor to recombination processes when neutral densities reach approximately $10^{-3}$ times the electron density ($n_e$).
Further details on the implementation of these processes are provided in Section \ref{subsection:charge-exchange}.

We use the lookup table method to pre-calculate the rate
coefficients of collisional ionisation using the analytical fit expression obtained by \citet{Voronov1997},  
radiative recombination coefficient fits from \citet{Zatsarinny_2003} and \citet{Kaur_2018}, and dielectronic recombination coefficient fits from \cite{Badnell_2006}.
During the evaluation of the source step, temperature and electron density are calculated and compared with the lookup table to find the closest temperature bins and electron density bins that are lower and higher than the cell values.
A linear interpolation is then performed to find an appropriate value for the collisional ionisation rate coefficient and radiative and dielectronic recombination coefficient.
Radiative cooling rates $\Lambda_{\kappa, i}(n_e, T)$ are
calculated using a bilinear interpolation method since they are functions of
both temperature and electron number density. We also generate a lookup table
for the mean photoionisation cross-section, utilizing the non-relativistic fitting formula by Verner et al. (1996), for species with ionisation threshold energies lower than the maximum energy of the energy bin array.

\begin{figure*}[ht]
\centering
\includegraphics[width=0.82\textwidth]{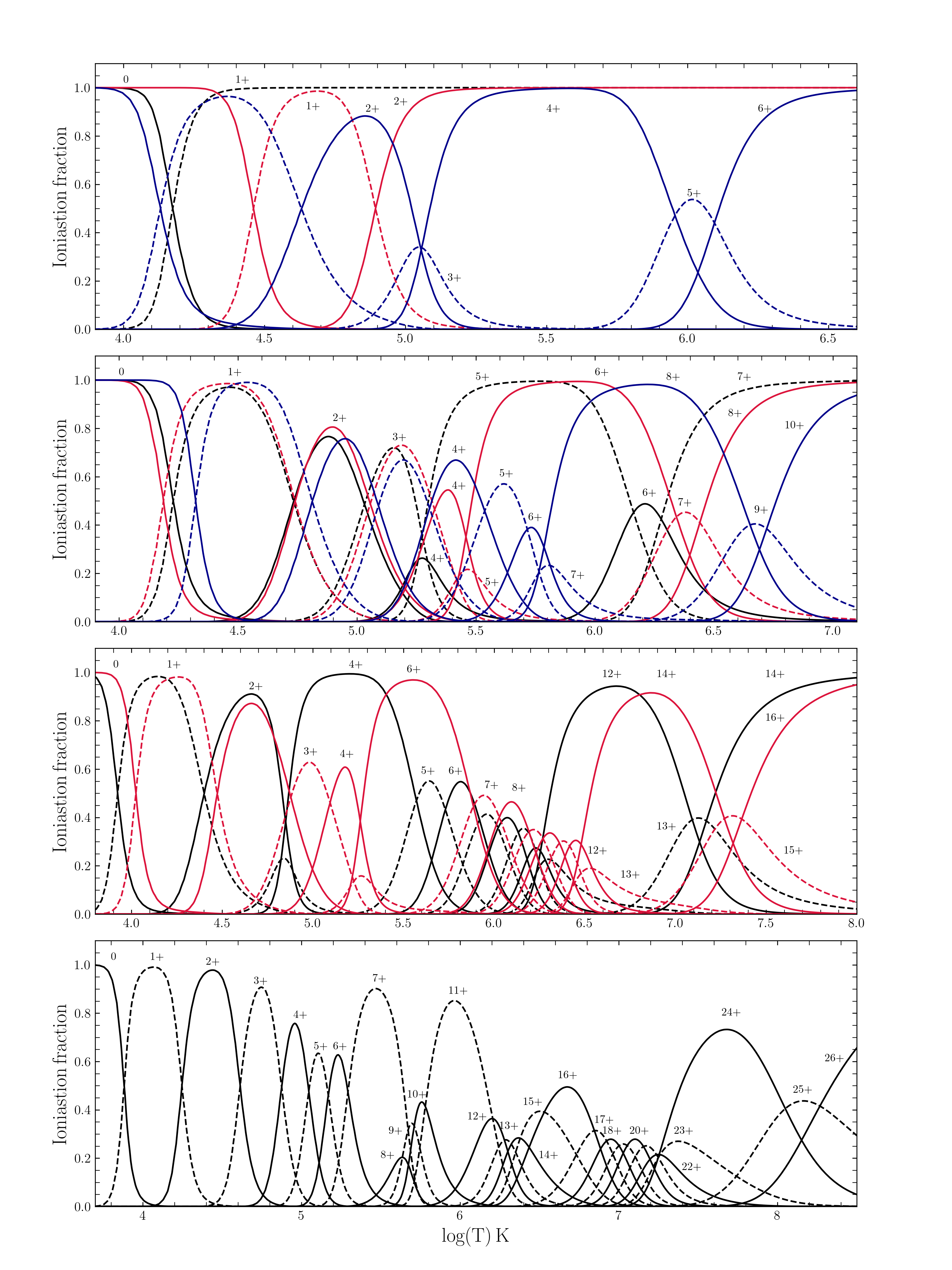}
\caption{ionisation fractions of the elements included in our microphysics module under CIE conditions, obtained by integrating until equilibrium is obtained.
The first panel depicts the ionisation levels of H (black), He (red), and C (blue); the second panel the ionisation levels of N (black), O (red), and Ne (blue); the third panel the ionisation levels of Si (black) and S (red); and the fourth panel the ionisation levels of Fe (black).}
\label{figure:CIE}
\end{figure*}

\subsection{Collisional ionisation and recombination}\label{subsection:ion-rec}
We use an empirical fit formula for electron collisional ionisation, such as the four-parameter fit formula adopted from \cite{Voronov1997}. This formula is accurate, with a deviation of less than 10\% from the recommended data, and avoids computationally expensive functions such as the exponential integral present in other available formulae \citep[e.g.,][]{Sutherland_1993}.

For radiative recombination, we utilized the analytical fit formula presented in \cite{Verner_1996}, incorporating a correction for low-charge ions as described in \cite{Badnell_2006}. This formula guarantees the accurate asymptotic behavior of rate coefficients for both low and high temperature regimes. The average accuracy of the fitting is not higher than the estimated absolute error, with smaller fitting errors for highly ionised species. This fit formula also agrees with power-law fits \cite{Pequignot_1991} and \cite{Arnaud_1992} in the temperature range in which they are valid, while covering a wider temperature range. However, this radiative recombination table does not contain all ions relevant to our calculation, specifically, the rates for Fe$^{+}$ through Fe$^{11+}$ were missing, hence we use old values from \cite{Arnaud_1992} that are refitted with Verner \& Ferland formula as described at \cite{Verner_1996}.

\begin{table}
\caption{The obtained Verner-Ferland fit parameters (A, B, T$_0$, T$_1$) for the S$^{1+}$ ion. These parameters were determined through the optimal fitting of the Verner-Ferland model onto the Aldrovandi and Pequignot fit with $A_{\text{rad}} = 4.1 \times 10^{-13}$ and $\eta = 0.630$.}
\begin{center}
\begin{tabular}{llll} 
 \hline
A & B &  T$_0$ &  T$_1$ \\
 \hline 
 $7.50 \times 10^{-12}$ & $6.45 \times 10^{-1}$ & $1.75 \times 10^{2}$ & $3.70 \times 10^{8}$ \\
\hline 
\end{tabular}
\end{center}
\label{table:VF_fit_parameters}
\end{table}

The literature lacks Verner-Ferland fit parameters for the radiative recombination of the S$^{1+}$ species. However, Aldrovandi and Pequignot have published fit parameters for this species \cite{Aldrovandi_1973}. We derived the Verner-Ferland fit parameters for S$^{1+}$ using the Aldrovandi-Pequignot fit parameters, $A_{\text{rad}} = 4.1 \times 10^{-13}$ and $\eta = 0.630$, by optimizing for the best fit. Our calculations yield the Verner-Ferland fit parameters as shown in Table \ref{table:VF_fit_parameters}.

We use dielectronic recombination rate coefficients detailed in \cite{Zatsarinny_2003, Badnell_2003}.
These coefficients are capable of providing accuracy to within 3\% for all ions in the temperature range of 10 to 10$^8$ K.
However, since this set does not include dielectronic recombination rate coefficients for S$^{1+}$, we have resorted to using the previous parameter set from \cite{Verner_1996} for this ion.

To model CIE conditions, we performed simulations of multiple zones with constant density of $n_\mathrm{H} = 1$\,cm$^{-3}$, in the absence of radiation.
Each zone was assigned an initial temperature between 10 K and $10^9$ K with step size of 0.0625 in $\log(T)$. We assume each zone was initially neutral, with the elemental mass-fractions set to Solar abundances according to \cite{Asplund_2009}, given in the second column of Table \ref{table:massfrac}.
Each zone was evolved adiabatically until the system achieved equilibrium.
Figure \ref{figure:CIE} displays the equilibrium ionisation balance as a function of temperature for all ions.
These can be compared with results in the literature, \citep[e.g.][]{Sutherland_1993}.

\begin{table}
\caption{Mass fractions for Solar abundances \citep{Asplund_2009} and for the the wind of a Wolf-Rayet star of type WC9 \citep[from][]{Eatson_2022}.}
\begin{center}
\begin{tabular}{ccc} 
 \hline
  Element & \multicolumn{2}{c}{X(E)} \\
          & Solar & WC9 \\      
\hline 
H  &  $0.738$ 				  & $0.0$\\ 
He &  $0.249$ 				  & $0.546$\\ 
C  &  $2.368 \times 10^{-3}$ & $0.4$\\ 
N  &  $6.935 \times 10^{-4}$ & $0.0$\\
O  &  $5.739 \times 10^{-3}$ & $0.05$\\ 
Ne &  $1.258 \times 10^{-3}$ & $1.258 \times 10^{-3}$\\
Si &  $6.656 \times 10^{-4}$ & $6.656 \times 10^{-4}$\\
S  &  $3.096 \times 10^{-4}$ & $3.096 \times 10^{-4}$\\
Fe &  $1.293 \times 10^{-3}$ & $1.293 \times 10^{-3}$	\\
\hline 
\end{tabular}
\end{center}
\label{table:massfrac}
\end{table}

\begin{figure}[ht]
\includegraphics[width=0.95\columnwidth]{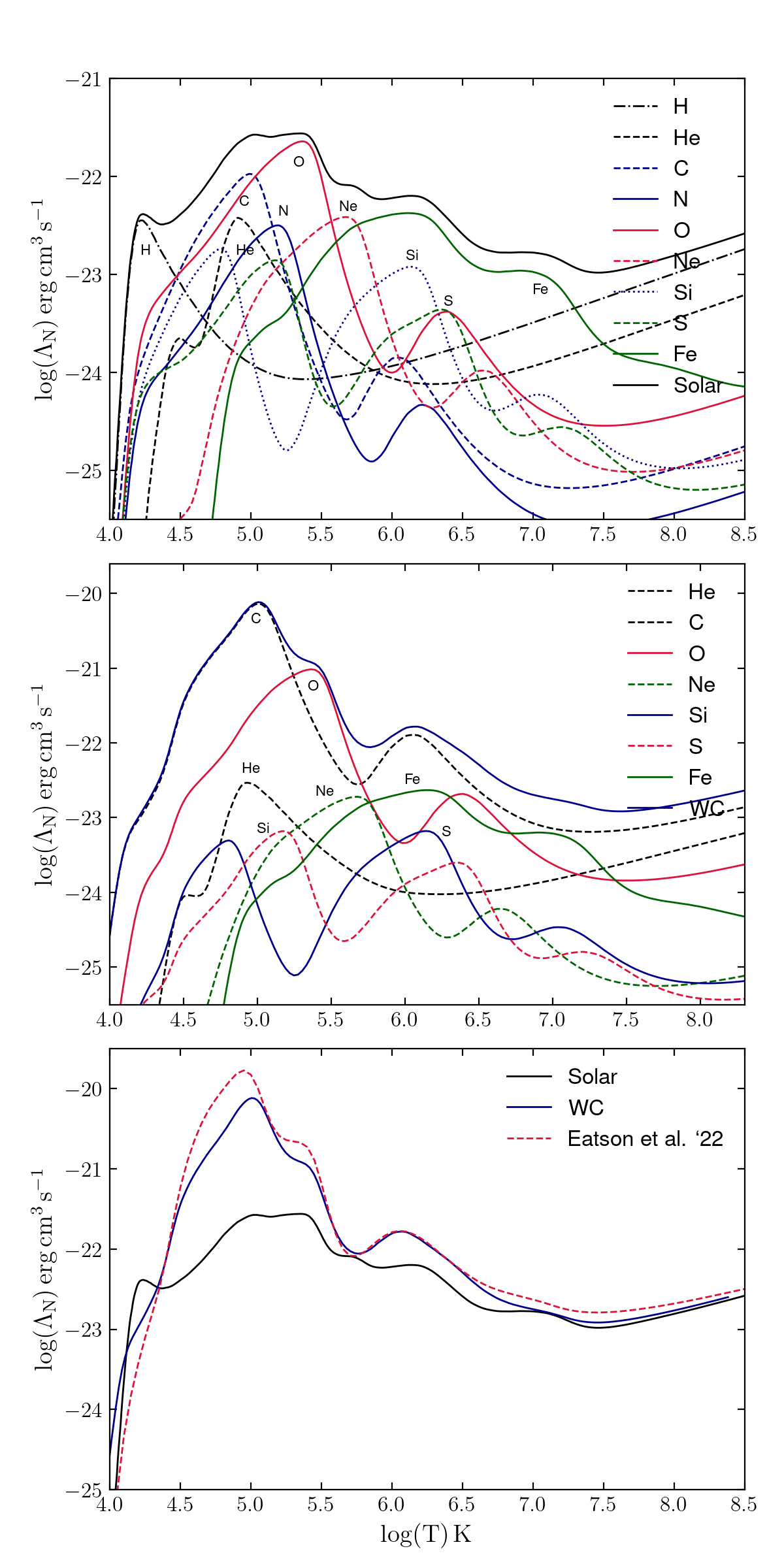}
\caption{Radiative cooling rates under CIE conditions obtained with our new microphysics module, including elemental contributions $L_{\kappa}$ (scaled with $(m_H/\rho)^2$) and net cooling function $\Lambda_{\rm N}$ for Solar and WC abundances.
Top: elemental contribution of the cooling curve for solar abundances, as specified in Table \ref{table:massfrac}. The black solid line represents the net cooling curve.
Middle: elemental contribution of the cooling curve for WC abundances, as indicated in Table \ref{table:massfrac}. The blue solid line represents the net cooling curve.
Bottom: net cooling curves for solar and WC abundances compared with the cooling function obtained by \citet{Eatson_2022}.}
\label{figure:cie_coolfn}
\end{figure}

\subsection{Radiative Cooling}\label{subsection:cie_cooling}

Cooling rates $\Lambda_{\kappa, i}(n_e, T)$ per electron, per ion of the species $\kappa$ with charge $i$ are calculated using the \textsc{CHIANTIPy} package v0.10.0 \citep{Zanna_2021, Young_2003}.
The radiative cooling calculation in \textsc{CHIANTIPy} accounts for the radiative loss resulting from the line-emissions of all spectral lines of the specific ion as a function of temperature and electron density~\citep{Young_2003, Dere_2009, Zanna_2015}.
In addition, bremsstrahlung emission is taken into account using the expression given in \cite{Rybicki_1986} with the Gaunt factor calculated using the method of \cite{Karzas_1961}. Furthermore, it includes two-photon emission of hydrogen \citep{Parpia_1982} and helium \citep{Drake_1986} isoelectronic sequence ions with the spectral distribution function taken from \cite{Goldman_1981} for hydrogen isoelectronic ions and \cite{Drake_1969} for helium isoelectronic ions, respectively.

The cooling rates, $\Lambda_{\kappa, i}(n_e, T)$, have been calculated and tabulated on an ion-by-ion basis in \texttt{NEMO}, without assuming ionisation equilibrium and for arbitrary abundances. 
These cooling rates are valid from $10\,\rm{K}$ to $10^9\,\rm{K}$ and electron densities ranging from $1\,\rm{cm}^{-3}$ to $10^6\,\rm{cm}^{-3}$.

To validate our implementation, we use the ion and electron number density under CIE conditions for $n_\mathrm{H}=1$\,cm$^{-3}$ and calculate the energy loss rate $L_{\kappa} = n_e\sum_i n_{\kappa, i} \Lambda_{\kappa, i}$ for each element.
The net energy loss rate, $L_{\rm cool}$, is obtained by summing over the individual species contributions. The normalized cooling function, given by 
\begin{equation}
\Lambda_{\rm N} = \left(\frac{m_H}{\rho}\right)^2 L_{\rm cool},
\end{equation}
is shown in Figure \ref{figure:cie_coolfn} as a function of $T$. The top panel in Figure \ref{figure:cie_coolfn} displays individual cooling function contributions of H, He, C, N, O, Ne, Si, S, and Fe under CIE conditions as a function of temperature,  assuming Solar abundances \citep{Asplund_2009}.
The net cooling rate exhibits several temperature peaks.
The initial peak at approximately $10^4$ K is primarily caused by Hydrogen Ly$\alpha$ cooling, and there are peaks at $10^5$ K, $4 \times 10^5$ K, $5 \times 10^5$ K, and $1.5 \times 10^6$ K, respectively, attributed to contributions from Carbon, Oxygen, Neon, and Iron. In addition, Iron gives a second peak at $2 \times 10^7$ K. 
Other elements also contribute to cooling, with Helium peaking at $4.9 \times 10^5$ K, Nitrogen at $5.1 \times 10^4$ K, and Silicon with two peaks at $6.3\times 10^4$ and $1.2\times 10^6$ K, respectively.
Sulfur also gives two peaks, at $1.5\times 10^5$ and $1.6\times 10^6$ K, respectively.
At higher temperatures, the cooling is dominated by thermal bremsstrahlung from fully ionised species.
These elemental CIE cooling results agree with previous computations, although differences in atomic data may lead to variances in detail, as has been observed in prior work \citep{Sutherland_1993, Wiersma_2009}.

The middle panel of Figure~\ref{figure:cie_coolfn} illustrates the cooling function for abundances appropriate for a WR star of type WC9, with the mass fractions of H, He, C, N, and O obtained from the study by \citet{Eatson_2022} and Solar abundances adopted for the remaining elements, given in the last column of Table~\ref{table:massfrac}.
Compared to the previous case of CIE with solar abundances, the net cooling rate for the WC9 abundance exhibits fewer peaks, with the primary peak occurring at around $10^5$ K due to carbon, followed by a second peak at $2.5 \times 10^5$ K due to oxygen and a third peak at $1.2 \times 10^6$ K due to carbon again.
The comparison of these two cooling curves in the bottom panel of Figure~\ref{figure:cie_coolfn} shows that the cooling for WC9 abundances is larger than for solar abundances. The cooling rate is about two orders of magnitude higher at around $10^5$ K and remains higher for $T>1.5\times 10^4$ K. However, in the temperature range $10^4<T<1.5\times 10^4$ K, cooling for solar abundances is higher.

\subsection{Photoionisation Implementation}\label{subsection:photoionisation}

As discussed above, we employ a variant of the C$^2$-ray method \citep{Mellema_2006} described by \citet{Mackey_2012} for photon-conserving radiative transfer and calculation of photoionisation rates,
using the method of short characteristics~\citep[e.g.][]{Raga_1999, Lim_2003} for raytracing.
In this update, we have extended the method to include several chemical species and multiple frequency bins for the radiation.
In this scheme, the photoionisation rate within a grid cell for the species ($\kappa, i$) to species ($\kappa, i+1$) is expressed as
\begin{equation}
\Gamma_{\kappa, i} = \int_{\nu^{\kappa, i}_{\rm th}}^{\infty} \frac{L_\nu}{h\nu} e^{-\tau_\nu} \frac{1- e^{-\Delta \tau_\nu^{\kappa, i}}}{n_{\kappa, i} V_{\rm shell}} \, d\nu \;\mathrm{s}^{-1}\,\mathrm{ion}^{-1} \;, 
\end{equation}
where $\tau_\nu$ is the optical depth from the source to the front edge of the cell, 
$\Delta \tau_\nu^{\kappa, i}$ represents the optical depth along a ray within the cell for the species ($\kappa, i$), $V_{\rm shell}$ has the same meaning as in \citet{Mellema_2006}, and $L_\nu$ is the spectral luminosity of the radiation source.

We further replace the frequency integration by Riemann summation over energy bins making the above equation
\begin{equation}
\Gamma_{\kappa, i} (r) = L_*\sum_{j = j_{\rm th}}^n F_j \left( \frac{e^{-\bar{\tau}_j}}{\bar{E}_j} \right) \frac{1- e^{-\Delta \bar{\tau}_j^{\kappa, i}}}{n_{\kappa, i} V_{\rm shell}} \;, 
\end{equation}
where the index $j$ iterates over the energy bins, $L_*$ represents the total luminosity of the star, and $F_j$ is the fraction of the total photon energy density in bin $j$. 
The average optical depth in each energy bin, denoted by $\bar{\tau}_j$, is computed by the ray-tracing module by summing cell optical depths along the ray. 
The cell optical depth for the species ($\kappa, i$) is given by
\begin{equation}
\Delta \bar{\tau}_j^{\kappa, i} = n_{\kappa, i} \bar{\sigma}_j^{\kappa, i} \Delta r \;,
\end{equation}
with $n_{\kappa, i}$ representing the number density of the photo-species ($\kappa, i$) in the cell, and $\Delta r$ the cell depth. Species with ionisation threshold energies lower than the maximum energy of the energy bins are treated as photo-species. 
The mean cross-section $\bar{\sigma}_j^{\kappa, i}$ for the species $(\kappa, i)$ in energy bin $j$ is determined following \citet{Mackey_2012} using the expression
\begin{equation}
\bar{\sigma}_j^{\kappa, i} = - \frac{1}{\sigma^{\kappa,i}(\bar{E}_j)} \ln \left( \int_j e^{-\sigma^{\kappa,i}(E_j)/\sigma^{\kappa,i}(\bar{E}_j)} \, dE_j \right),
\end{equation}
where $\bar{E}_j$ is the mean energy of bin $j$. The integration is evaluated using the composite Simpson's 3/8 rule over the range of energy bin $j$.
The photoionisation cross sections, $\sigma^{\kappa,i}(E)$, in the above equation are calculated using the analytic fit formula presented in \cite{Verner_1996b}.

Photo heating $Q_{\rm photo}$ is calculated in a similar manner 
and is given by
\begin{equation}
Q_{\rm photo} = \sum_{\kappa, i} n_{\kappa, i} Q_{\kappa, i}.
\end{equation}
The summation is performed over photo-species, and the individual contributions $Q_{\kappa, i}$ are calculated following \citet{Mellema_2006} using the equation
\begin{equation}
Q_{\kappa, i} = L_*\sum_{j = j_{\rm th}}^n F_j \left( \frac{e^{-\bar{\tau}_j}}{\bar{E}_j} \right) (\bar{E}_j - E_{\rm th}^{\kappa, i}) \frac{1- e^{-\Delta \bar{\tau}_j^{\kappa, i}}}{n_{\kappa, i} V_{\rm shell}}.
\end{equation}

The binned spectral energy distribution, $F_j$ in the above expressions are pre-calculated for ATLAS stellar atmosphere models \citep{Castelli_2004} and Potsdam Wolf-Rayet models \citep{Hamann_2004, Sander_2012,Todt_2015}.
The ATLAS9 model library encompasses a wide range of elemental abundances, represented by $\rm[M/H]$, spanning from solar metallicity ($0.0$) to metal-poor conditions ($-2.5$).
Furthermore, it covers a broad gravity range from $\log g = 0.0$ to $+5.0$, with increments of $+0.5$.
The effective temperature grid ranges from 3500 K to 50000 K, but it is not evenly spaced.
These models are designed to provide spectral information spanning from $1.6\times10^6$\,\AA \ to 90.9 \AA, with varying wavelength intervals.
The Potsdam Wolf-Rayet Models gives spectral profiles of Wolf-Rayet stars of the nitrogen subclass (WN). 
They offer two distinct varieties: hydrogen-free models (WNE) and models with a specified mass fraction of hydrogen (WNL). 
Moreover, these models cater to diverse metallicities, corresponding to the iron-group and total CNO mass fractions, which align with the metallicities of the Galaxy, the Large Magellanic Cloud (LMC), or the Small Magellanic Cloud (SMC), respectively.

For calculations in this paper the radiation field was divided into 15 energy bins in the range $h\nu\in[7.9, 77.0]$\,eV.
The boundaries of the energy bins were chosen to coincide with the most important ionisation energies for the ions considered, and are quoted in Table~\ref{tab:bin-energy}.
The raytracing module in \textsc{pion} only tracks direct radiation from sources to the point it is absorbed, and does not track scattered or re-emitted radiation.
Instead we assume the On-the-Spot (OTS) approximation, where recombinations to the ground state of H emit a photon that is assumed to locally reionise another H atom.
To achieve agreement between  \textsc{pion} and \textsc{Cloudy} for the radius of the He$^+$ and H$^+$ ionised zones (see section~\ref{subsection:hiiregion}), it was necessary to correctly account for re-ionisations of H and He by recombinations of He$^{2+}$ and He$^+$.
We followed the method of \citet{FriMelIli12} to correctly allocate recombination photons to the ionisation of H and He.
The algorithm will be considerably more complicated to implement for arbitrary elemental abundances; this is deferred to future work.
To calculate the recombinations into different energy levels the case A recombination rate is augmented with case B recombination rates for H$^+$, He$^+$ and He$^{2+}$, taking the fitting functions from \citet{HuiGne97}.

\subsection{Charge-exchange}\label{subsection:charge-exchange}
\begin{figure*}
\centering
\includegraphics[width=0.8\textwidth]{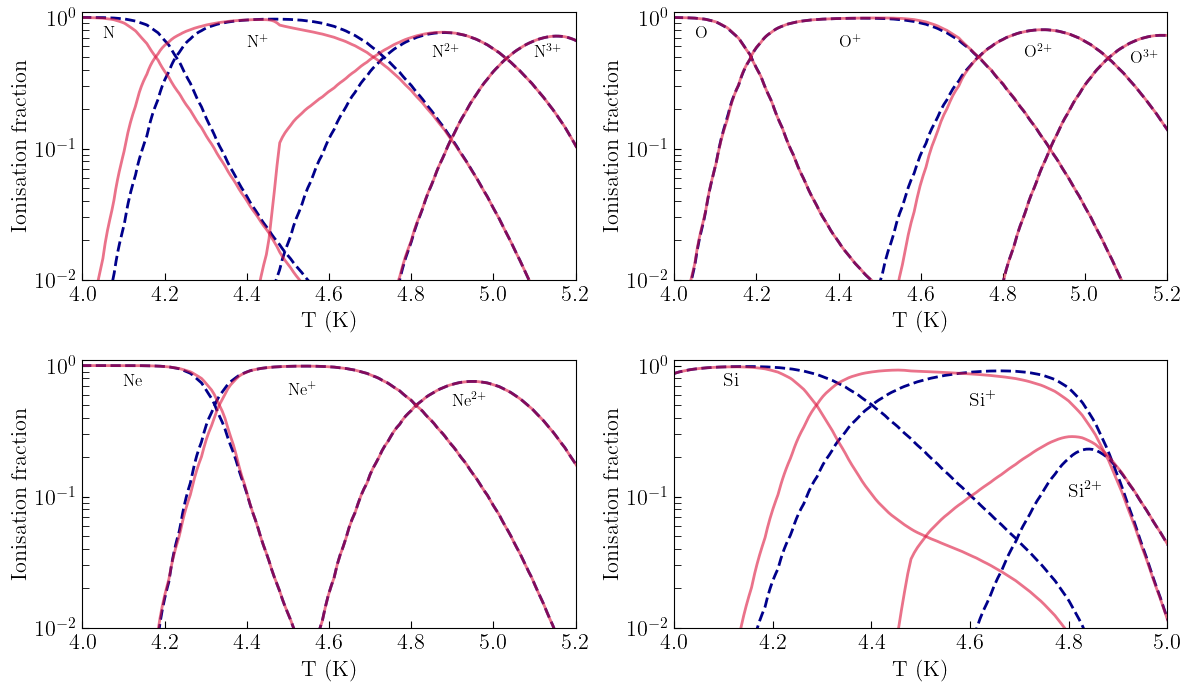}
\caption{Comparison of CIE balance for N, O, Ne, and Si incorporating and omitting charge exchange reactions with H and He. Solid curves represent ionisation fractions with charge exchange calculations, while dashed curves represent ionisation fractions without charge exchange reactions.}
\label{figure:CX_CIE}
\end{figure*}

\begin{table}
\caption{List of recombination charge exchange reactions with their respective energy defects. The last column provides the reference index for the energy defect values. References: (1) \cite{Zygelman1990}, (2) \cite{Friedman2017}, (3) \cite{ButlerDalgarno1980}, (4) \cite{ButlerDalgarno1979}, (5) \cite{ButlerRaymond1980}, (6) \cite{McCarrolVairon1976}, (7) \cite{NeufeldDalgarno1987}, (8) \cite{Steigman1975}}
\begin{tabular}{lll}
\hline
Reaction & $\Delta E$ (eV) & Ref. \\
\hline
 \multicolumn{3}{c}{Recombination reaction with H} \\
\hline
 $\mathrm{He}^{+} + \mathrm{H} \rightarrow \mathrm{He} + \mathrm{H}^{+}$ & $-8.83$ & (1) \\
$\mathrm{He}^{2+} + \mathrm{H} \rightarrow \mathrm{He}^{1+} + \mathrm{H}^{+}$   & $40.82$ &  (2) \\

$\mathrm{C}^{+} + \mathrm{H} \rightarrow \mathrm{C} + \mathrm{H}^{+}$   & $-2.33$ & (3) \\
$\mathrm{C}^{2+} + \mathrm{H} \rightarrow \mathrm{C}^{+} + \mathrm{H}^{+}$   & $10.78$ & (2) \\
$\mathrm{C}^{3+} + \mathrm{H} \rightarrow \mathrm{C}^{2+} + \mathrm{H}^{+}$   & $21.60$ & (2) \\
$\mathrm{C}^{4+} + \mathrm{H} \rightarrow \mathrm{C}^{3+} + \mathrm{H}^{+}$   & $13.34$ & (2) \\

$\mathrm{N}^{+} + \mathrm{H} \rightarrow \mathrm{N} + \mathrm{H}^{+}$ &   $0.94$ & (4) \\
$\mathrm{N}^{2+} + \mathrm{H} \rightarrow \mathrm{N}^{+}  + \mathrm{H}^{+}$ &   $11.95$ & (2) \\
$\mathrm{N}^{3+} + \mathrm{H} \rightarrow \mathrm{N}^{2+}  + \mathrm{H}^{+}$ &   $0.00$ & --- \\
$\mathrm{N}^{4+} + \mathrm{H} \rightarrow \mathrm{N}^{3+}  + \mathrm{H}^{+}$ &   $0.00$ & --- \\

 $\mathrm{O}^{+} + \mathrm{H} \rightarrow \mathrm{O} + \mathrm{H}^{+}$ &   $0.02$ & (5) \\
 $\mathrm{O}^{2+} + \mathrm{H} \rightarrow \mathrm{O}^{+} + \mathrm{H}^{+}$ &   $16.50$ &  (2)\\
 $\mathrm{O}^{3+} + \mathrm{H} \rightarrow \mathrm{O}^{2+} + \mathrm{H}^{+}$ &   $0.00$ & --- \\
 $\mathrm{O}^{4+} + \mathrm{H} \rightarrow \mathrm{O}^{3+} + \mathrm{H}^{+}$ &   $0.00$ & ---\\

$\mathrm{Ne}^{2+} + \mathrm{H} \rightarrow \mathrm{Ne}^{+} + \mathrm{H}^{+}$ & $0.00$ &  ---\\
$\mathrm{Ne}^{3+} + \mathrm{H} \rightarrow \mathrm{Ne}^{+} + \mathrm{H}^{+}$ & $0.00$ &  ---\\
$\mathrm{Ne}^{4+} + \mathrm{H} \rightarrow \mathrm{Ne}^{+} + \mathrm{H}^{+}$ & $8.60$ &  (2)\\
 
$\mathrm{Si}^{2+} + \mathrm{H} \rightarrow \mathrm{Si}^{+} + \mathrm{H}^{+}$ & $2.74$ &  (6)\\
$\mathrm{Si}^{3+} + \mathrm{H} \rightarrow \mathrm{Si}^{2+} + \mathrm{H}^{+}$ & $0.00$ &  ---\\
$\mathrm{Si}^{4+} + \mathrm{H} \rightarrow \mathrm{Si}^{3+} + \mathrm{H}^{+}$ & $0.00$ &  ---\\   

$\mathrm{S}^{1+} + \mathrm{H} \rightarrow \mathrm{S} + \mathrm{H}^{+}$ &   $-3.24$ &  (3)\\
$\mathrm{S}^{2+} + \mathrm{H} \rightarrow \mathrm{S}^{+} + \mathrm{H}^{+}$ &  $0.00$ &  ---\\
$\mathrm{S}^{3+} + \mathrm{H} \rightarrow \mathrm{S}^{2+} + \mathrm{H}^{+}$ &  $0.00$ &  ---\\
$\mathrm{S}^{4+} + \mathrm{H} \rightarrow \mathrm{S}^{3+} + \mathrm{H}^{+}$ & $0.00$ &  ---\\

$\mathrm{Fe}^{2+} + \mathrm{H} \rightarrow \mathrm{Fe}^{+} + \mathrm{H}^{+}$ &   $2.56$ &  (7)\\
$\mathrm{Fe}^{3+} + \mathrm{H} \rightarrow \mathrm{Fe}^{2+} + \mathrm{H}^{+}$ & $0.00$ &  ---\\
\hline
\hline
\multicolumn{3}{c}{Recombination reaction with He} \\
\hline
$\mathrm{C}^{2+} + \mathrm{He} \rightarrow \mathrm{C}^+ + \mathrm{He}^{+}$ & $-0.2$ & (8) \\
$\mathrm{C}^{3+} + \mathrm{He} \rightarrow \mathrm{C}^{2+} + \mathrm{He}^{+}$ & $16.81$ & (2) \\
$\mathrm{C}^{4+} + \mathrm{He} \rightarrow \mathrm{C}^{3+} + \mathrm{He}^{+}$ & $31.89$ & (2) \\

$\mathrm{N}^{2+} + \mathrm{He} \rightarrow \mathrm{N}^+ + \mathrm{He}^{+}$ & $0.00$ & --- \\
$\mathrm{N}^{3+} + \mathrm{He} \rightarrow \mathrm{N}^{2+} + \mathrm{He}^{+}$ & $0.00$ & ---\\
$\mathrm{N}^{4+} + \mathrm{He} \rightarrow \mathrm{N}^{3+} + \mathrm{He}^{+}$ & $0.00$ & --- \\

$\mathrm{O}^{2+} + \mathrm{He} \rightarrow \mathrm{O}^+ + \mathrm{He}^{+}$ & $10.52$ & (2) \\
$\mathrm{O}^{3+} + \mathrm{He} \rightarrow \mathrm{O}^{2+} + \mathrm{He}^{+}$ & $25.00$ & (2) \\
$\mathrm{O}^{4+} + \mathrm{He} \rightarrow \mathrm{O}^{3+} + \mathrm{He}^{+}$ & $0.00$ & --- \\

$\mathrm{Ne}^{2+} + \mathrm{He} \rightarrow \mathrm{Ne}^+ + \mathrm{He}^{+}$ & $16.37$ & (2) \\
$\mathrm{Ne}^{3+} + \mathrm{He} \rightarrow \mathrm{Ne}^{2+} + \mathrm{He}^{+}$ & $0.00$ & --- \\
$\mathrm{Ne}^{4+} + \mathrm{He} \rightarrow \mathrm{Ne}^{3+} + \mathrm{He}^{+}$ & $0.00$ & --- \\

$\mathrm{Si}^{3+} + \mathrm{He} \rightarrow \mathrm{Si}^{2+} + \mathrm{He}^{+}$ & $8.88$ &  \\
$\mathrm{Si}^{4+} + \mathrm{He} \rightarrow \mathrm{Si}^{3+} + \mathrm{He}^{+}$ & $0.00$ & --- \\

$\mathrm{S}^{3+} + \mathrm{He} \rightarrow \mathrm{S}^{2+} + \mathrm{He}^{+}$ & $0.00$ & --- \\
$\mathrm{S}^{4+} + \mathrm{He} \rightarrow \mathrm{S}^{3+} + \mathrm{He}^{+}$ & $0.00$ & --- \\
\hline
\end{tabular}
\label{table:recomb_charge_exchange}
\end{table}

Charge exchange can be an important process in partially ionised zones where H (or He) is mostly neutral but other species may be ionised (e.g.\ O).
We consider recombination reactions with H and He of the type:
\begin{equation}
    B^{q} + A \rightarrow B^{(q-1)} + A^+ + \Delta E,
\end{equation}
where $q$ is set from 1 to 4, $A$ is H or He, and $B^q$ is the recombining ion with charge $q$. The reverse ionisation reactions involve H$^+$ and He$^+$ donating an electron to ionise a heavier ion:
\begin{equation}
B^{q-1} + A^+ \rightarrow B^{q} + A + \Delta E.
\end{equation}
The energy defect $\Delta E$ in the above equation can be either positive or negative, depending on the reaction.

We base our calculations on the parameters provided by \cite{KingdonFerland1996} for the analytic fit formula of \citet{ArnaudRothenflug1985}.
This formula is used to determine the rate coefficient for recombination charge-exchange reactions with H, as listed in Table \ref{table:recomb_charge_exchange}.
The corresponding rate coefficients for reverse (ionisation) reactions are listed in Table \ref{table:ionise_charge_exchange} and are related to the forward rates by the principle of detailed balance, expressing the ionisation rate coefficients as a product of the fitting function and the Boltzmann factor $\exp(- \Delta E/kT)$. 
For He charge-transfer recombination reactions, also shown in Table \ref{table:recomb_charge_exchange}, we apply the fitting formula and parameters from \cite{ArnaudRothenflug1985}. 
We generate a lookup table for the ionisation and recombination reaction rate coefficients $k_{\kappa,i}^{s,q}(T)$, pertaining to the reactant species $(\kappa, i)$ and $(s,q)$ as a function of temperature.

The source term due to charge exchange can be explicitly expressed as:
\begin{multline}\label{eq:cx_source}
    \tilde{S}_{\kappa,i}^{\rm cx} = Y_{\kappa,i+1} \sum_{s,q} n_{s,q} \, k_{\kappa, i+1}^{s,q} - 2 Y_{\kappa,i} \sum_{s,q} n_{s,q} \, k_{\kappa, i}^{s,q} \\ 
    + Y_{\kappa,i-1} \sum_{s,q} n_{s,q} \, k_{\kappa, i-1}^{s,q} 
\end{multline}
The first term in this expression represents the contribution of all reactions leading to the recombination of species transitioning from \((\kappa, i+1) \rightarrow (\kappa, i)\). The last term accounts for the ionisation of species transitioning from \((\kappa, i-1) \rightarrow (\kappa, i)\). 
The second term addresses both recombination and ionisation, representing the transitions of species from \((\kappa, i) \rightarrow (\kappa, i-1)\) and \((\kappa, i) \rightarrow (\kappa, i+1)\), respectively.
These transitions, driven by charge transfer, can heat the system as shown in Tables \ref{table:recomb_charge_exchange} and \ref{table:ionise_charge_exchange} which indicate that most reactions are exothermic. 
The heating due to charge transfer (\(Q_{\rm cx}\)) is given by:
\begin{equation}
    Q_{\rm cx} = \sum_{\kappa, i}  n_{\kappa, i} \sum_{s,q} n_{s,q} \, \Delta E_{\kappa, i}^{s,q} \, k_{\kappa, i}^{s,q}, 
\end{equation}
where \(\Delta E_{\kappa, i}^{s,q}\) is the energy gain from the reaction involving the reactant species \((\kappa, i)\) and \((s,q)\).

\begin{table}
\caption{List of ionisation charge exchange reactions with their respective energy defects. References: (1) \cite{ButlerRaymond1980}, (2) \cite{ButlerHeilDalgarno1980} }
\begin{tabular}{lll}
\hline
Reaction & $\Delta E$ (eV) & Ref. \\
\hline
 \multicolumn{3}{c}{Ionisation reaction with H$^+$} \\
\hline
$\mathrm{C} + \mathrm{H}^+ \rightarrow \mathrm{C}^+ + \mathrm{H}$ & $2.33$ & Inverse Reaction Value \\
$\mathrm{N} + \mathrm{H}^+ \rightarrow \mathrm{N}^{+} + \mathrm{H}$ & $-0.94$ & Inverse Reaction Value \\
$\mathrm{O} + \mathrm{H}^+ \rightarrow \mathrm{O}^{+} + \mathrm{H}$ & $-0.02$ & Inverse Reaction Value \\
$\mathrm{Si}^{+} + \mathrm{H}^+ \rightarrow \mathrm{Si}^{2+} + \mathrm{H}$ & $-2.74$ & (1) \\
$\mathrm{S}^{+} + \mathrm{H}^+ \rightarrow \mathrm{S}^{+} + \mathrm{H}$ & $0.00$ & --- \\
$\mathrm{Fe}^{+} + \mathrm{H}^+ \rightarrow \mathrm{Fe}^{2+} + \mathrm{H}$ & $0.00$ & --- \\
\hline
\hline
\multicolumn{3}{c}{Ionisation reaction with He$^+$} \\
\hline
$\mathrm{C}^{+} + \mathrm{He}^+ \rightarrow \mathrm{C}^{2+} + \mathrm{He}$ & $-6.29$ & (2) \\
$\mathrm{N}^{+} + \mathrm{He}^+ \rightarrow \mathrm{N}^{2+} + \mathrm{He}$ & $0.00$ & --- \\
$\mathrm{Si}^{+} + \mathrm{He}^+ \rightarrow \mathrm{Si}^{2+} + \mathrm{He}$ & $0.00$ & --- \\

$\mathrm{Si}^{2+} + \mathrm{He}^+ \rightarrow \mathrm{Si}^{3+} + \mathrm{He}$ & $-8.88$ & (1) \\
$\mathrm{S}^{+} + \mathrm{He}^+ \rightarrow \mathrm{S}^{2+} + \mathrm{He}$ & $0.00$ &  --- \\
$\mathrm{S}^{2+} + \mathrm{He}^+ \rightarrow \mathrm{S}^{3+} + \mathrm{He}$ & $0.00$ &  --- \\
\hline
\end{tabular}
\label{table:ionise_charge_exchange}
\end{table}

To confirm the precision of calculation, we 
conduct collisional ionisation equilibrium simulations using the same initial conditions as our previous one reported in Section \ref{subsection:ion-rec}, but with charge exchange reactions enabled and \( q \) set to 4.
Figure \ref{figure:CX_CIE} illustrates the equilibrium ionisation balance for N, O, Ne, and Si, comparing scenarios with and without charge exchange reactions over the temperature range of \( 10^4-10^5 \) K.
The ionisation profiles exhibit significant differences, aligning well with the results from the literature \citep{Gu2022, ArnaudRothenflug1985}.
The slight variations in the profiles are due to minor changes in the rate coefficient values.
 
\section{Results}\label{section:results}

Here we examine a series of 1D calculations using \texttt{NEMO} for adiabatic shocks, radiative shocks, and H~\textsc{ii} regions that can be compared with solutions from the literature.
We first examine shock models similar to Model E from \cite{Raymond1979} within a 1D Cartesian framework, specifically employing slab symmetry. 
The setup features a uniform inflow from the positive $x$-boundary and a reflecting condition at the negative $x$-boundary.
Table~\ref{tab:shocks} outlines the simulation domain alongside the uniform initial density $\rho_0$, velocity $v_{x,0}$, temperature $T_0$, and magnetic field $B_{y,0}$. 
The other velocity and magnetic field components are zero in the initial conditions and remain zero due to the symmetry.

Model E in \cite{Raymond1979} utilizes abundance set A, which is based on the cosmic abundances from \cite{Allen1973}.
We adopt the same abundance set, with the exception of Mg and Ca. Consequently, we track 99 chemical species across the four shock tests detailed in the following subsections.
These tests aim to assess the accuracy of our calculations concerning the distinct flow properties and physical processes presented in Table~\ref{tab:shocks}.

The H~\textsc{ii} region tests are based on the HII40 Lexington benchmark test \citep{Fer95}, including multi-frequency ionising radiation, all ionisation/recombination  and heating/cooling processes, but no hydrodynamics.

\begin{table*}
 \centering
  \caption{Properties of the shock tests described in sections~\ref{subsection:adia_shock} to \ref{subsection:cx_shock}.  Column meanings are: $L_\mathrm{box}$ is the length of the $x$-domain. $N_\mathrm{cells}$ is the number of grid cells spanning this domain (uniform grid), $\Delta x=L_\mathrm{box}/N_\mathrm{cells}$, $\rho_0$, $v_{x,0}$, $T_0$ and $B_{y,0}$ are the uniform initial conditions, and the last two columns indicate whether or not radiative cooling and charge-transfer (CT) reactions are included.} 
  \label{tab:shocks}
  \begin{tabular}{@{}l l l l l l l l l l}
  \hline
   Test/Parameter  &  $L_\mathrm{box}$ & $N_\mathrm{cells}$  & $\Delta x$  & $\rho_0$  &  $v_{x,0}$  & $T_0$ & $B_{y,0}$  & Cooling & CT\\
    (Unit) &  ($10^{16}$ cm) &  & ($10^{12}$ cm) & ($10^{-23}$\,g\,cm$^{-3}$) &   (km\,s$^{-1}$) & (K) &  (G) &  &\\
 \hline
    Shock Test 1a &  5.0  & 1024  & 4.8828E+1 & 2.0374 &  -100  & 1.0E+4 & 1.0E$-6$ & No & No\\
    Shock Test 1b &  5.0  & 10240  & 4.8828E+0 & 2.0374 &  -100  & 1.0E+4 & 1.0E$-6$ & No & No\\
    
    Shock Test 2 &  10$^4$  & 1536 & 6.5104E+4 & 5.0935 &  -1000  & 1.0E+4 & 1.0E$-6$ & Yes & No\\

    Shock Test 3a &  1.0  & 10240  & 9.7660E$-1$ & 5.0935 &  -1000
    & 1.0E+4 & 1.0E$-6$ & Yes & No\\
    Shock Test 3b &  1.0  & 10240  & 9.7660E$-1$ & 5.0935 &  -3000  & 1.0E+4 & 1.0E$-6$ & Yes & No\\

    Shock Test 4 &  0.3  & 256  & 1.1719E+1 & 2.0374 &  -120 & 1.0E+4 & 1.0E$-6$ & Yes & Yes\\
\hline
\end{tabular}
\end{table*}

\subsection{Adiabatic shock test 1}\label{subsection:adia_shock}

\begin{figure}
\includegraphics[width=1\columnwidth]{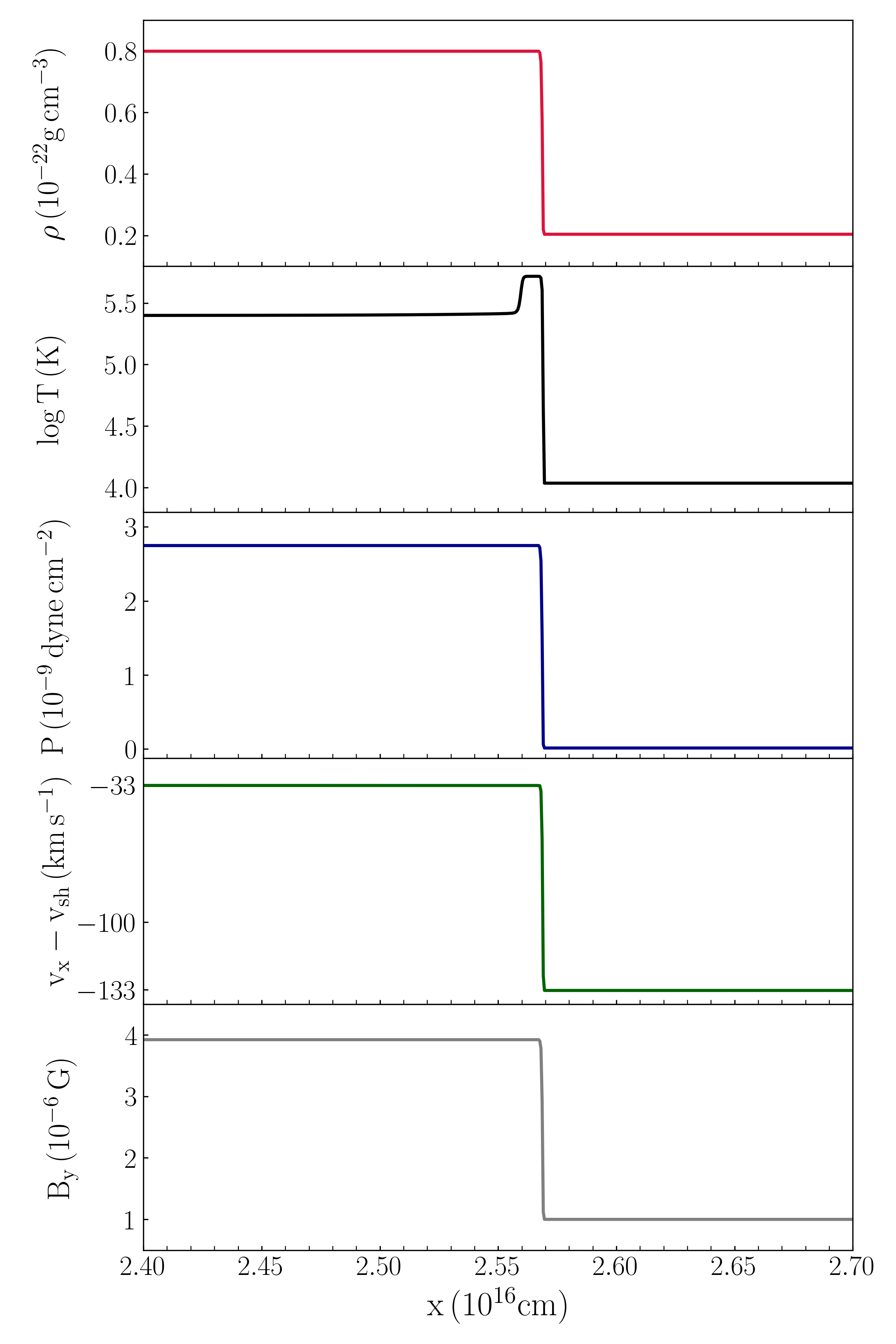}
\caption{Flow quantities of an adiabatic shock generated by gas flowing at a velocity of 100\,km\,s$^{-1}$. The simulation employs 10240 grid cells and not all of the domain is shown.
}
\label{Fig_HL_resolution_FlowStructure}
\end{figure}

Shock Test 1a is an adiabatic shock with Mach number of about 10.
Figure \ref{Fig_HL_resolution_FlowStructure} illustrates the flow structure of the shock, which is formed by the inflowing neutral gas travelling at a speed of 100 km s$^{-1}$ with a pre-shock particle number density $n$ of 10 cm$^{-3}$. The shock velocity, $v_\mathrm{sh}=133$\,km\,s$^{-1}$, is larger than 100\,km\,s$^{-1}$ because it reflects off the negative boundary and propagates upstream at velocity  $(v_\mathrm{sh}/4)$\,km\,s$^{-1}$ by conservation of mass in the (adiabatic) downstream layer.
The Rankine-Hugoniot jump condition compresses the gas by a factor of 4 and predicts a post-shock temperature of
\begin{equation}
T = \frac{3}{16} \frac{\mu m_H}{k_B} v_\mathrm{sh}^2 \;,
\end{equation}
where $\mu$ is the mean mass per particle in unit of the hydrogen atom mass, $m_H$, and we assume an adiabatic index of $\gamma=5/3$ for the gas equation of state.
For $v_\mathrm{sh}=133$\,km\,s$^{-1}$ this results in post-shock temperature of $T=4.9\times10^5$\,K.

The kinetic energy of the gas upstream of the shock is converted into thermal energy as it transitions through the shock front, resulting in an increase in the temperature of the gas downstream, and consequently collisional ionisation.

In the postshock relaxation layer, this thermal energy ionises the gas by removing ionising potential energy from the gas, as given by the expression
\begin{equation}
\varepsilon = \frac{9}{8} \rho_0 v_\mathrm{sh}^2 - \sum_{\kappa, i} n_{\kappa, i} I_{\kappa, i},
\end{equation}
where $I_{\kappa, i}$ denote the ionisation potential of the ion of the species $\kappa$ and charge $i$. Due to the rapid ionisation, we observe an drop in the temperature to $2.5\times10^{5}$ K immediately after the shock.
This is almost entirely due to H ionisation because it is the most abundant element.
In this way the neutral state of the pre-shock gas significantly influences the post-shock temperature far downstream from the shock.

\begin{figure*}
\centering
\includegraphics[width=0.9\textwidth]{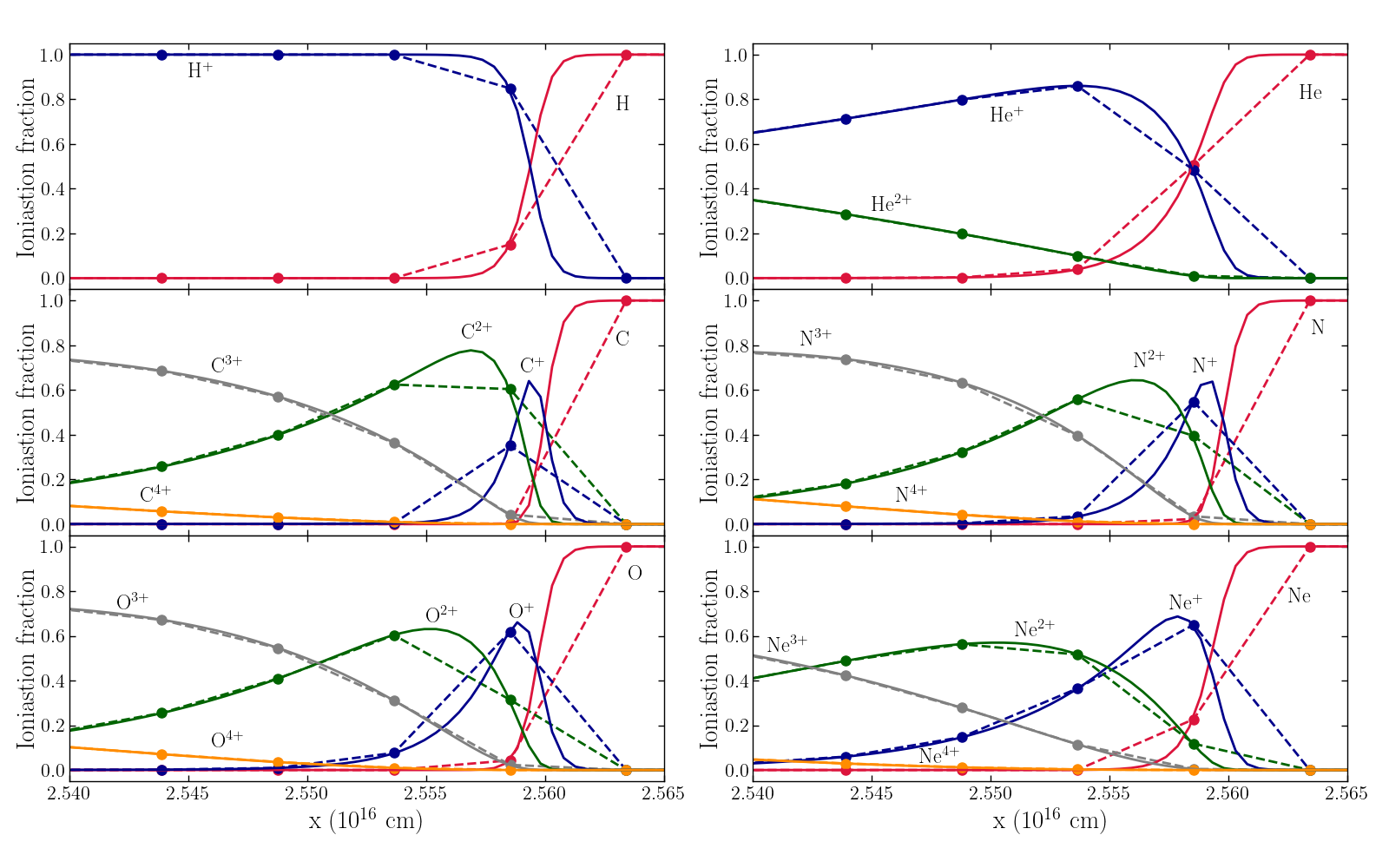} 
\caption{Spatial profiles of ion fractions for H, He, C, N, O, and Ne across the shock front for shock test 1, comparing the low (dashed lines) and high (solid lines) resolution simulations. The flow moves from right to left, with the shock positioned at the plot's right-hand boundary. The plot depicts only the downstream region.}
\label{Fig_HL_resolution_HHeCN0Ne}
\end{figure*}

Figure \ref{Fig_HL_resolution_HHeCN0Ne} depicts the ionisation fraction profiles across the shock front region for a gas composed of neutral atoms of H, He, C, N, O, Ne, Si, S, and Fe.
In this context, where radiative losses are absent, the gas tends towards equilibrium to the left, stabilizing at a temperature of $2.51\times10^{5}$ K. 
At this equilibrium temperature, hydrogen and helium demonstrate nearly complete ionisation, with ionisation fractions of 0.999 for H$^{1+}$ and 0.998 for He$^{2+}$. Among other elements, notable ionisation states include C$^{4+}$ at 0.989, N$^{4+}$ at 0.104, N$^{5+}$ at 0.866, O$^{3+}$ at 0.224, O$^{4+}$ at 0.533, O$^{5+}$ at 0.144, Ne$^{3+}$ at 0.214, Ne$^{4+}$ at 0.664, Ne$^{5+}$ at 0.111, Si$^{4+}$ at 0.914, S$^{6+}$ at 0.904, Fe$^{6+}$ at 0.129, and Fe$^{7+}$ at 0.857. These values closely match the ionisation fraction obtained in the CIE test depicted in Figure~\ref{figure:CIE} for a temperature of $2.51\times10^{5}$ K. The gas achieves the ionisation state over a finite timescale, determined by the specific recombination and ionisation rates of the species at the given temperature.

Figure \ref{Fig_HL_resolution_HHeCN0Ne} compares the ionisation profiles of H, He, C, N, O, Ne, Si, S, and Fe across the shock front for both low (1024 grid points, $\Delta x = 4.8828\times10^{13}$ cm) and high resolution (10240 grid points, $\Delta x = 4.8828\times10^{12}$ cm) simulations (shock test 1b) for the same domain size of $5.0\times10^{16}$ cm. Despite the substantial resolution discrepancy, these simulations exhibit noteworthy agreement, particularly away from the shock. However, this agreement diminishes in the immediate vicinity of the shock, primarily due to the rapid temperature variation across the shock. As the gas swiftly ionises to a state consistent with the local temperature, facilitated by the small ionisation timescale, significant spatial changes occur near the shock front, necessitating a higher resolution grid to accurately capture these nuances. Yet, this requirement pertains specifically to the immediate shock region, with overall agreement between low and high resolution simulations remaining relatively robust.

\subsection{Non-adiabatic shock tests 2 and 3}\label{subsection:nonadia_shock}
\begin{figure}
\includegraphics[width=\columnwidth]{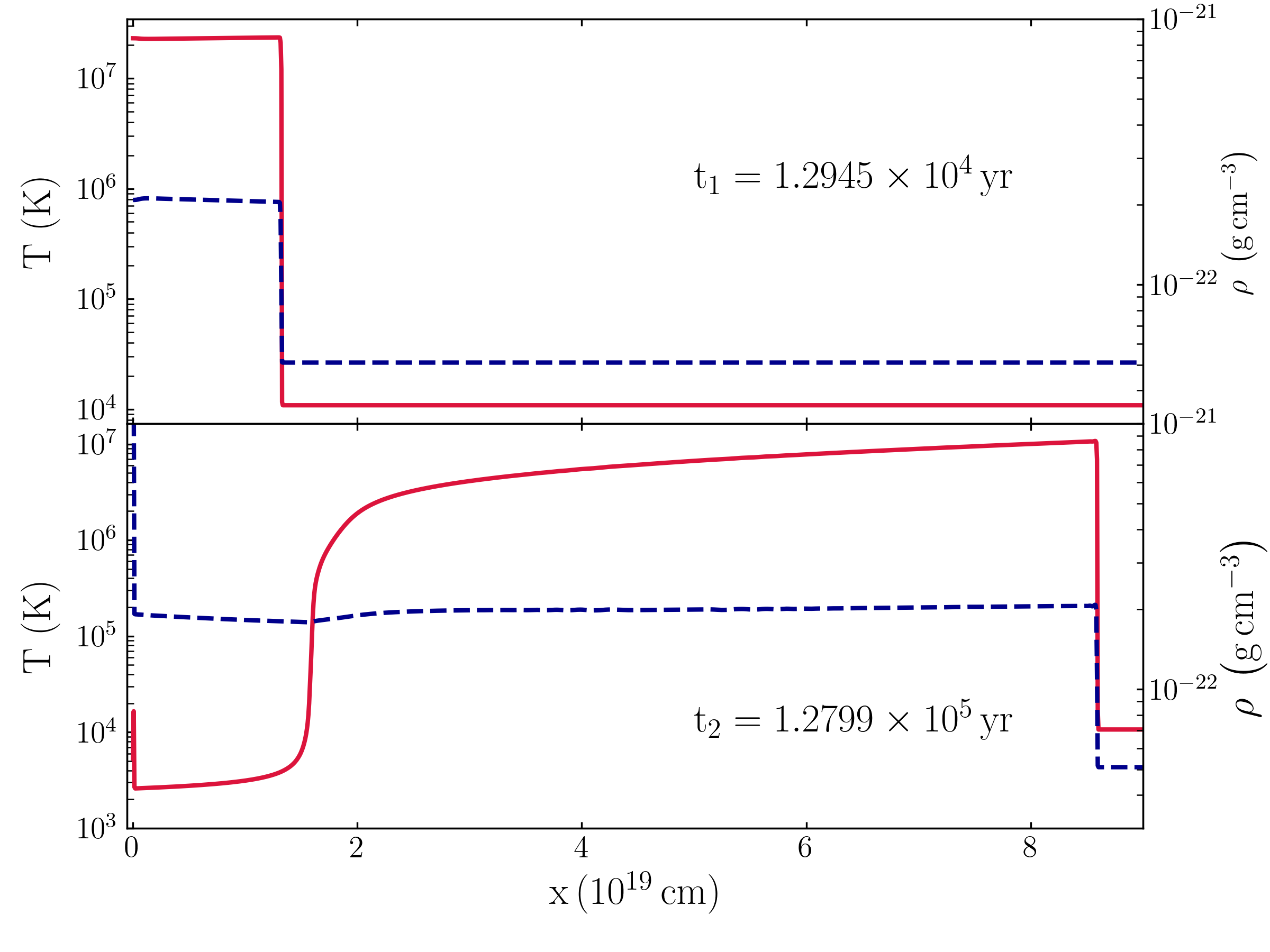}
\caption{Illustration of the flow structure in pre-shock gas interacting with the shock front. The top panel exhibits temperature and density profiles at $t_1=1.2945\times10^{4}$ years, while the bottom panel presents the same at a later time, $t=1.2799\times10^{5}$ years, highlighting the interplay between Chianti cooling data and gas dynamics.}
\label{figure:non-adiabatic_flow}
\end{figure}

\begin{figure}
\includegraphics[width=\columnwidth]{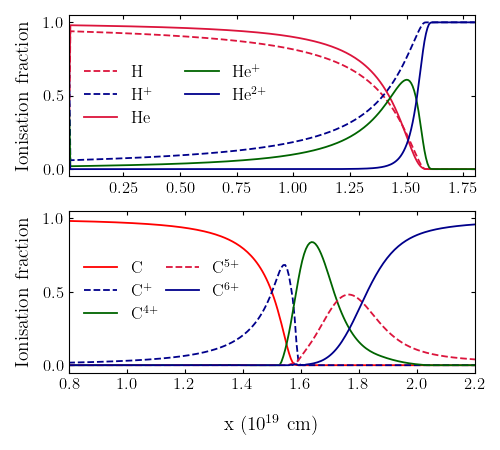}
\caption{Ionisation fractions of H, He, and C within the cooling region of non-adiabatic shock test 2, revealing non-equilibrium dynamics of chemical species at later time, $t=1.2799\times10^{5}$ years.}
\label{figure:non-adiabatic_ionisation}
\end{figure}

As an extension of the preceding planar shock test, we conduct the shock tests under non-adiabatic conditions for higher inflow velocities with cooling turned on and no radiation sources.
The flow structure of a neutral pre-shock gas of density $5.0935 \times 10^{-23}$ g/cm$^3$ intersecting the shock front at a velocity of 1000 km/s are illustrated in Figure \ref{figure:non-adiabatic_flow}.
The two panels display temperature and density profiles at two distinct time points.
Initially, at $t_1=1.2945\times10^{4}$ years, the post-shock gas remains uncooled, maintaining a temperature of $T\sim1.5\times10^{7}$ K, determined by the Rankine–Hugoniot jump conditions, because the postshock cooling timescale is longer than the simulation has been running for.
Later, at $t_2=1.2799\times10^{5}$ years, the downstream gas, particularly the region farthest from the shock front, undergoes significant cooling, with the temperature dropping to $T\sim3.0\times10^{3}$ K, as depicted in the bottom panel of Figure \ref{figure:non-adiabatic_flow}.

The top (H and He) and bottom (C) panels of the Figure \ref{figure:non-adiabatic_ionisation} show the non-equilibrium ionisation structure of of the post-shock gas as it undergoes cooling.
Initially,
H, He, and C exist in a fully ionised state immediately behind the shock front.
As the downstream gas gradually cools from an initial temperature of $T\sim1.5\times10^{7}$ K to $T\sim3.0\times10^{3}$ K, one can observe the transition in the ionisation state of the gas, progressing from its most ionised state to a neutral state (from right to left).

\begin{figure*}
\centering
\includegraphics[width=\textwidth]{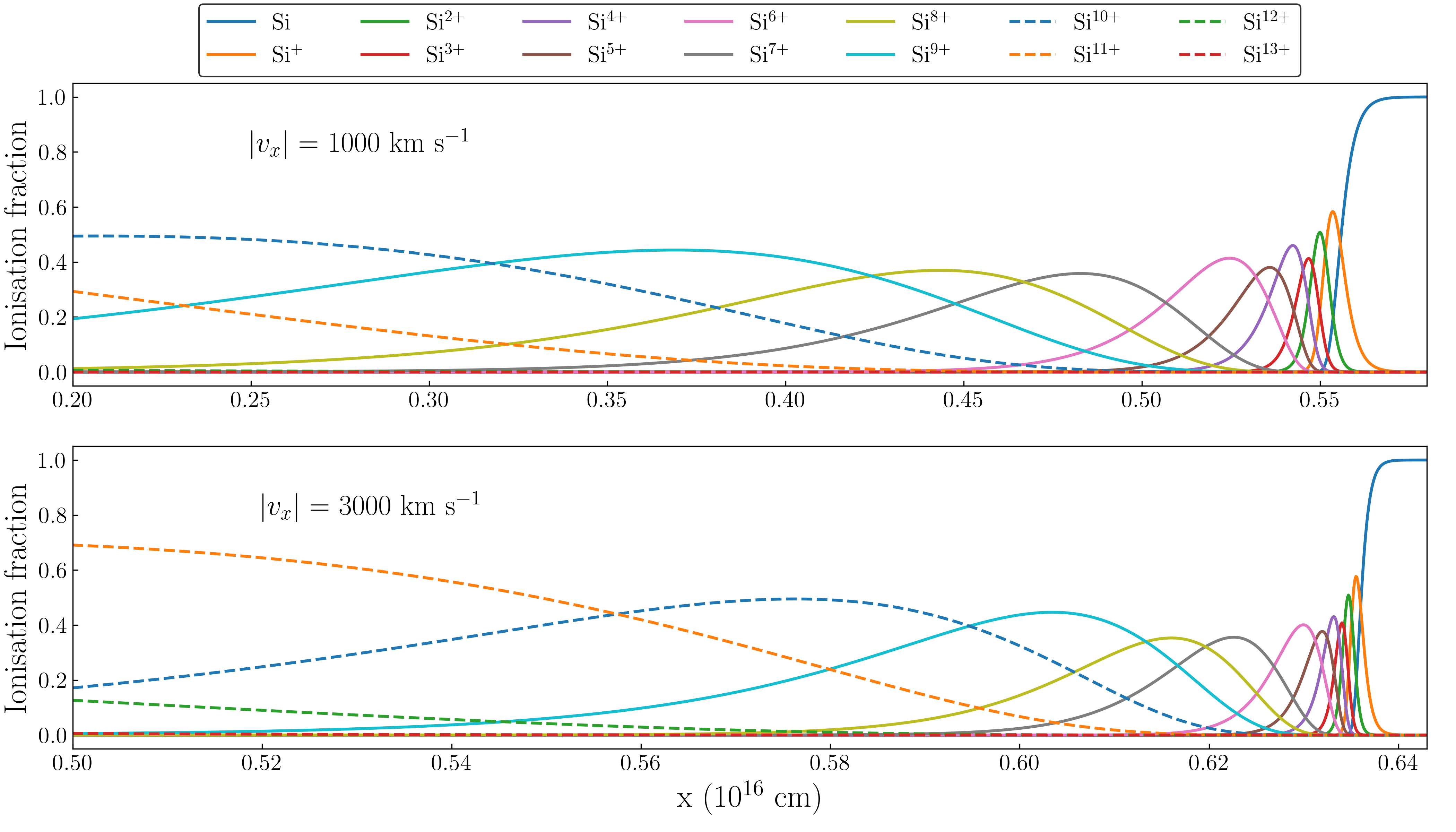}
\caption{Spatial profiles of ion fractions for different ionisation levels of Si across the shock front in non-adiabatic flows at velocities of 1000 km/s (top panel) and 3000 km/s (bottom panel).}
\label{figure:nonadia_compare_Si}
\end{figure*}

Figure \ref{figure:nonadia_compare_Si} presents a comparison of the ionisation profiles of Si for two higher inflow velocities: 1000 km/s in the top panel (shock test 3a) and 3000 km/s in the bottom panel (shock test 3b).
Here the postshock temperature is sufficiently high that the cooling lengthscale is larger than the simulation domain and the shock behaves adiabatically.
However, the ionisation profiles in the immediate vicinity of the shock front show a clear dependence on the inflow velocity.
 
The higher post-shock temperature associated with larger inflow velocity of 3000 km s$^{-1}$ results in increased ionisation rates, subsequently leading to shorter ionisation timescales.
This phenomenon, as discussed in the previous section, implies that smaller ionisation levels of chemical species exhibit reduced spatial extents.
To resolve the spatial profile of ionisation fractions for each ionisation state of the elements 
within the post-shock ionisation layer, a fine spatial resolution is imperative.
Consequently, the simulations were conducted with 10240 grid cells to ensure precise spatial delineation.

The ionisation process unfolds progressively from right to left, initiating from the shock front. This sequential progression occurs due to the finite ionisation time scale associated with each ionisation state. Behind the shock, the temperature rises sufficiently to enable ionisation across multiple levels of Si ions. The significance of higher ionisation states manifests only at a specific spatial distance from the shock front, owing to the time required for successive ionisation processes, contingent upon the gas temperature in that region being high enough to facilitate such ionisation. Consequently, the emission spectra observed immediately behind the shock primarily reflect the presence of lower ionisation states within the gas chemical species. Conversely, emission from highly ionised gas is observable away from the shock front.

\subsection{Charge transfer Non-adiabatic shock test 4}\label{subsection:cx_shock}

In this section, we extend the previous non-adiabatic shock model to incorporate charge exchange with \(q=4\) (see Section \ref{subsection:charge-exchange}), tracing all ionisation levels of H, He, C, N, O, Ne, Si, S, and Fe. This integration includes all the reactions listed in Tables \ref{table:recomb_charge_exchange} and \ref{table:ionise_charge_exchange}. As in previous cases, we assume a neutral preshock gas, but now with a slower inflow velocity of 120 km/s. This adjustment results in a steady-state shock similar to the one presented in \cite{ButlerRaymond1980}.

\begin{figure}
\includegraphics[width=\columnwidth]{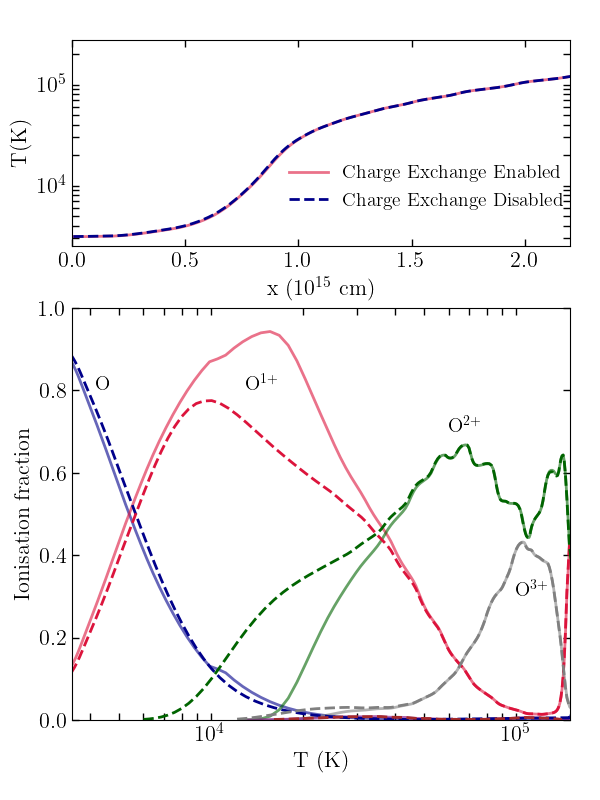}
\caption{Temperature and ionisation profiles of the shocked gas with and without charge transfer processes. The top panel shows temperature as a function of position, with the shock off the grid to the right, and the flow from right to left.  The temperature profile with and without charge-exchange is almost identical. The bottom panel shows the ionisation fractions of oxygen, emphasizing the significance of charge transfer reactions for the oxygen ionisation state, especially below $6 \times 10^3$ K.}
\label{figure:cx_temp_oxygen}
\end{figure}

Figure \ref{figure:cx_temp_oxygen} compares the temperature distribution and ionisation states of O behind the shock front with and without charge exchange reactions after $\sim 5\times10^{2}$ yrs. The minimal variation in the temperature profile indicates that heating due to the charge exchange reactions listed in Tables \ref{table:recomb_charge_exchange} and \ref{table:ionise_charge_exchange} is insignificant. However, there is a notable impact on the ionisation of the gas, as shown by the ionisation states of O in the bottom panel of Figure \ref{figure:cx_temp_oxygen}. 

When the temperature decreases to \(10^5\) K, the reaction rate for \(\mathrm{O}^{2+} + \mathrm{H} \rightarrow \mathrm{O}^{+} + \mathrm{H}^{+}\) becomes significant, with a value around \(\sim10^{-9}\) cm$^3$\,s$^{-1}$, considerably higher than the corresponding radiative recombination rate, although at this high temperature the electron abundance is much larger than the neutral H abundance, reducing the importance of charge exchange. At temperature \(30\,000\) K, the fractional abundance of neutral hydrogen increases enough to allow the charge-exchange reaction to become the dominant process. At \(20\,000\) K, the $\mathrm{O}^{+}/\mathrm{O}^{2+}$ ratio is significantly larger compared to the case when charge exchange reactions are omitted.
Below \(10\,000\) K, the reaction \(\mathrm{O}^{+} + \mathrm{H} \rightarrow \mathrm{O} + \mathrm{H}^{+}\) gains importance, maintaining a rate of about \(\sim10^{-9}\) cm$^3$\,s$^{-1}$.
The reverse ionisation reaction \(\mathrm{O} + \mathrm{H}^{+} \rightarrow \mathrm{O}^{+} + \mathrm{H}\) also proceeds rapidly, with a rate \(\sim9 \times 10^{-10}\) cm$^3$\,s$^{-1}$.
These competing reactions ensure ionisation balance remains largely unaffected by charge transfer.

\subsection{HII Region test}\label{subsection:hiiregion}

\begin{figure}
\centering
\includegraphics[width=0.9\columnwidth]{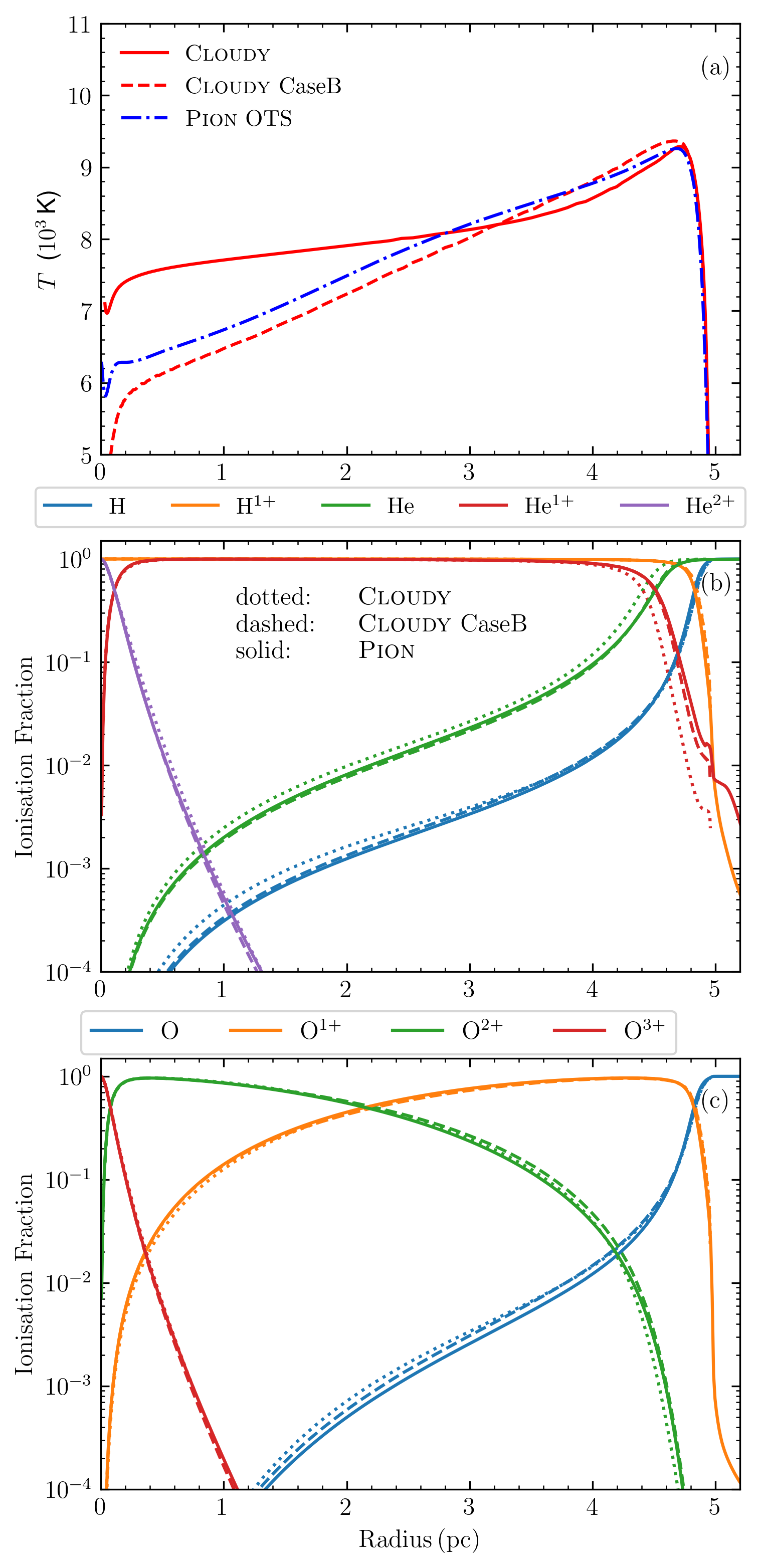}
\caption{Photoionisation equilibrium calculation for the temperature and ionisation state of a static H~\textsc{ii} region for a H number density $n_\mathrm{H}=10\,\mathrm{cm}^{-3}$:
(a) gas temperature vs.\ radius for 3 different calculations: solid-red line obtained with \textsc{Cloudy}, dashed red line \textsc{Cloudy} with the CaseB setting and dot-dashed blue line shows the result with \textsc{pion} using the On-the-Spot approximation (see text for details);
(b) ionisation state of H and He as a function of radius: solid lines are results obtained with \textsc{pion}, dotted lines with \textsc{Cloudy} and dashed lines \textsc{Cloudy}+CaseB; and
(c) as (b) but showing ionisation state of O as a function of radius.}
\label{fig:hiiregion-n1}
\end{figure}

\begin{figure}
\centering
\includegraphics[width=0.9\columnwidth]{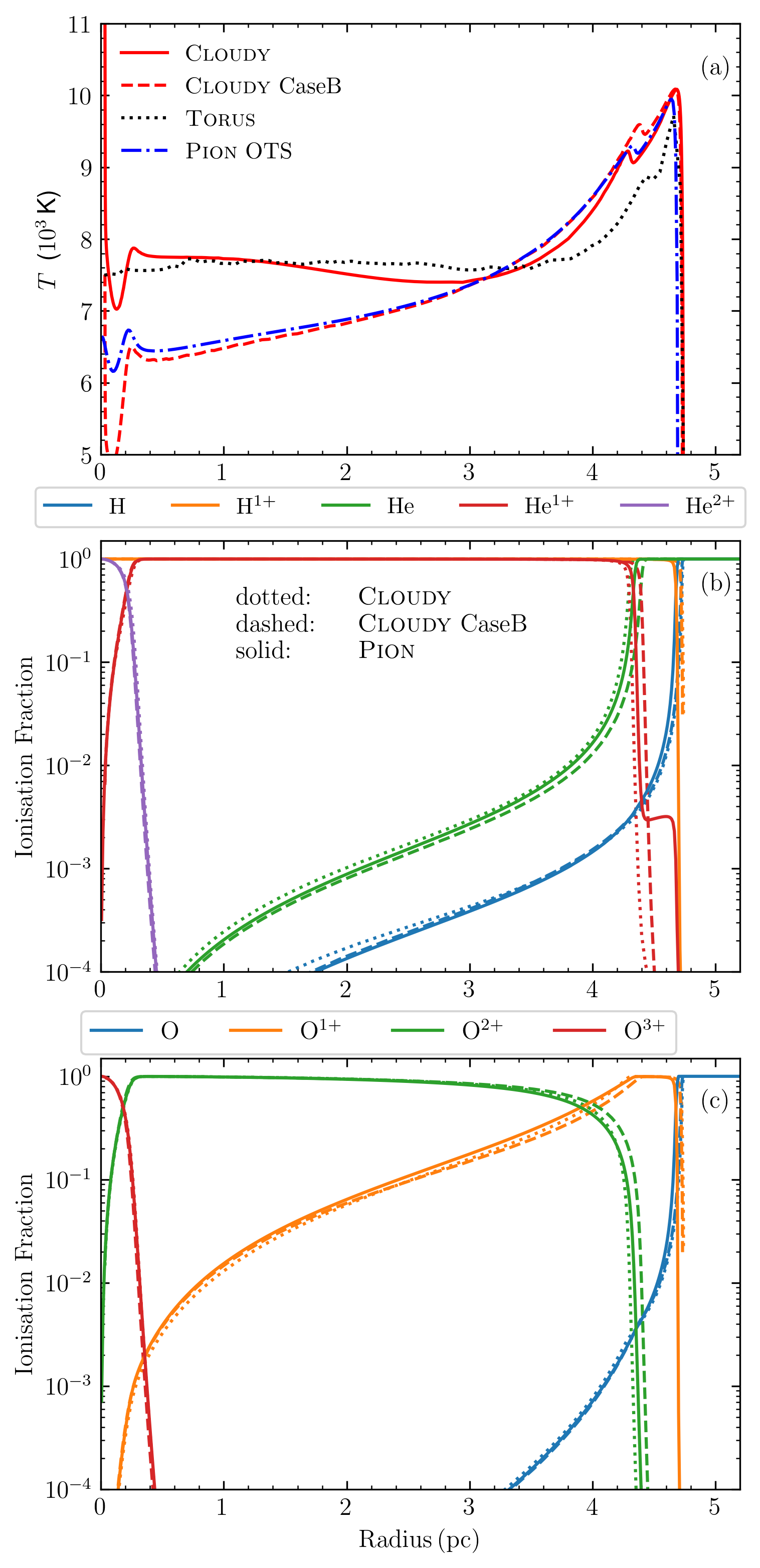}
\caption{Photoionisation equilibrium calculation for the temperature and ionisation state of a static H~\textsc{ii} region for a H number density $n_\mathrm{H}=100\,\mathrm{cm}^{-3}$ (The Lexington HII40 benchmark test): panels (a), (b), and (c) as for Fig.~\ref{fig:hiiregion-n1} except that panel (a) also includes the result from \textsc{Torus}.}
\label{fig:hiiregion-n2}
\end{figure}

\begin{figure}
\centering
\includegraphics[width=0.9\columnwidth]{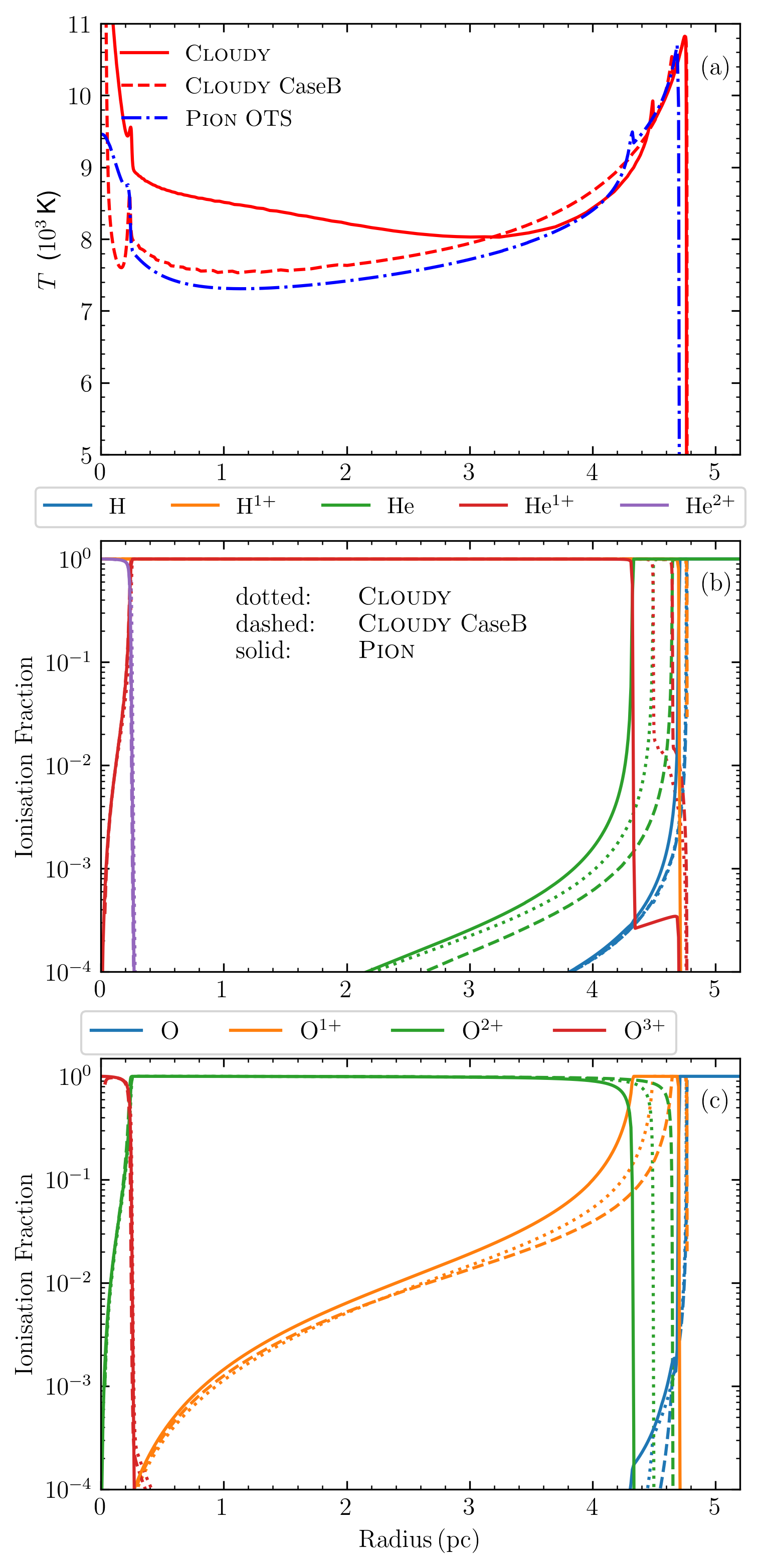}
\caption{Photoionisation equilibrium calculation for the temperature and ionisation state of a static H~\textsc{ii} region for a H number density $n_\mathrm{H}=1000\,\mathrm{cm}^{-3}$: panels (a), (b), and (c) as for Fig.~\ref{fig:hiiregion-n1}.}
\label{fig:hiiregion-n3}
\end{figure}

\begin{table}
 \centering
  \caption{Simulation parameters used for the H\,\textsc{ii} region test with the HII40 Lexington benchmark model \citep{Fer95, ErcBarSto03, HawHar12}.
  Test 2 is the HII40 model, and Tests 1 and 3 have gas density 10 times lower and higher, respectively, with adjustments to the radiation field such that the STr\"omgren radius remains the same.
  Note in the benchmark test Si and Fe are not included and so we exclude them from our calculations.}
  \label{tab:HIIregion}
  \begin{tabular}{@{}l l l l}
   \textbf{Stellar parameters (Unit)} & \textbf{Test 1} & \textbf{Test 2} & \textbf{Test 3} \\
 \hline
   $\textrm{T}_{\rm{eff}} \textrm{(K)}$ & \multicolumn{3}{c}{40\,000} \\
   Spectral energy distribution & \multicolumn{3}{c}{Blackbody}  \\
   ${R}_*(R_\odot)$  & 1.865 & 18.65 & 186.5 \\
   $L_*$ ($10^{39}$\,erg\,s$^{-1}$)     & 0.0307 & 3.07 & 307.0 \\
   $Q_0$ ($10^{49}$\,s$^{-1}$) & 0.0426 & 4.26 & 426.0 \\
     \hline\hline
   \textbf{ISM properties (Unit)} & \textbf{Test 1} & \textbf{Test 2} & \textbf{Test 3}  \\
 \hline
   $\rho$  ($10^{-22}$\,g\,cm$^{-3}$) & 0.236 & 2.36 & 23.6 \\
   X$_{\rm H}$  & \multicolumn{3}{c}{$7.0933\times 10^{-01}$} \\
   X$_{\rm He}$ & \multicolumn{3}{c}{$2.8373\times 10^{-01}$} \\
   X$_{\rm C}$  & \multicolumn{3}{c}{$1.8725\times 10^{-03}$} \\
   X$_{\rm N}$  & \multicolumn{3}{c}{$3.9726\times 10^{-04}$}  \\
   X$_{\rm O}$  & \multicolumn{3}{c}{$3.7452\times 10^{-03}$}  \\
   X$_{\rm Ne}$ & \multicolumn{3}{c}{$7.1647\times 10^{-04}$} \\
   X$_{\rm Si}$  & \multicolumn{3}{c}{$1.9932\times 10^{-09}$} \\ 
   X$_{\rm S}$  & \multicolumn{3}{c}{0} \\ 
   X$_{\rm Fe}$  & \multicolumn{3}{c}{0} \\ 
   Domain size (cm) & \multicolumn{3}{c}{$1.8\times 10^{19}$} \\
   $n_{\rm{cells}}$ & \multicolumn{3}{c}{$256$} \\
   Finish time ($10^{13}$\,s) & 31.56 & 3.156 & 0.3156 \\
   \hline
\end{tabular}
\end{table}

\begin{table}
 \centering
  \caption{Energy bins for the radiation field for photoionisation calculations.
  The bins are chosen to line up with ionisation thresholds of the modelled ions, as much as possible.
  }
  \label{tab:bin-energy}
  \begin{tabular}{@{}l l l}
  \hline
  Bin & $E_\mathrm{min}$ (eV) & $E_\mathrm{max}$ (eV) \\
  \hline
  0 & 7.9 & 11.3 \\
  1 & 11.3 & 13.6 \\
  2 & 13.6 & 14.5 \\
  3 & 14.5 & 16.2 \\
  4 & 16.2 & 21.6 \\
  5 & 21.6 & 24.4 \\
  6 & 24.4 & 29.6 \\
  7 & 29.6 & 30.7 \\
  8 & 30.7 & 35.1 \\
  9 & 35.1 & 41.0 \\
  10 & 41.0 & 45.1 \\
  11 & 45.1 & 47.9 \\
  12 & 47.9 & 54.4 \\
  13 & 54.4 & 63.4 \\
  14 & 63.4 & 77.0 \\
  \hline
\end{tabular}
\end{table}

To validate the photo-ionisation subroutine, we conducted an HII region simulation following the HII40 Lexington benchmark test \citep{Fer95}.
The simulation utilized a one-dimensional, uniform grid with spherical symmetry, comprising 512 grid zones and covered a radial domain extending from the origin to $r=1.8\times 10^{19}$ cm.
At the center is a blackbody radiation source with an effective surface temperature of $40\,000$\,K, irradiating an initially neutral, cold and uniform ISM.
Hydrodynamics is switched off, and the system is integrated until photoionisation equilibrium is achieved.
Three calculations were run, with parameters as indicated in Table~\ref{tab:HIIregion}: Test 2 is the HII40 Lexington benchmark model \citep{Fer95} (note that the elemental abundances are quoted in terms of mass fraction rather than the more typical number fraction relative to H) and Tests 1 and 3 are variants with lower and higher gas density, respectively.
The ionising photon output of the source was decreased/increased for tests 1 and 3 to give the same Str\"omgren radius as for Test 2 (test 3 obviously has an unreasonably large luminosity for a stellar source and is run only for comparison with the other two tests).

We used \textsc{Cloudy} version C23.01 \citep{ChaBiaCam23} to make a comparison with our results.
For Test 2 we also compared the results with published data from \textsc{Torus} \citep{HawHar12}.
After finding differences between the results from \textsc{pion} and \textsc{Cloudy} for the temperature structure of the H~\textsc{ii} region, we also ran \textsc{Cloudy} with the CaseB setting, which effectively switches off transport of recombination radiation beyond the Lyman limit by setting the local optical depth for this radiation to very large values.
This imposes the well-known OTS approximation (see section~\ref{subsection:photoionisation}), where recombinations to the ground state are assumed to result in local ionisations.
This assumption is not correct for some gas densities, but it more closely mimics the radiative transfer in \textsc{Pion}, which makes the OTS approximation.

Results for Tests 1, 2 and 3 are shown in Figs.~\ref{fig:hiiregion-n1},~\ref{fig:hiiregion-n2} and~\ref{fig:hiiregion-n3}, respectively.
In each case the top panel shows gas temperature as a function of radius for the converged final state; the middle panel shows ionisation fractions of H and He; and the bottom panel ionisation fractions of ions of O.
The main result shown here is the close agreement between \textsc{Pion} and \textsc{Cloudy} in `Case B' mode for the temperature and ionisation fraction as a function of radius, especially for tests 1 and 2.
The location of the He$^{2+}$, He$^+$ and H$^+$ ionisation fronts are in good agreement; only for the high-density test 3 is the location of the He$^+$ ionisation front offset significantly.
The ionisation state of O follows closely that of He and H, and so the two codes show good agreement for tests 1 and 2, but diverge somewhat for test 3 at the $\textrm{O}^{2+}\rightarrow\textrm{O}^+$ transition.
For test 3 there is also a significant difference in the location of the He$^+$ ionisation front between \textsc{Cloudy} runs with and without Case B, and so we suspect this is related to the simplified radiative transfer treatment in \textsc{Pion}.

The reason for the divergent temperature structure between \textsc{Pion} and the standard settings of \textsc{Cloudy} appears to be in the transport of the diffuse radiation field generated by recombinations.
\textsc{Pion} uses a simple raytracer to transport radiation directly from the source until it is absorbed, without scattering, and assumes any re-emitted or scattered radiation is re-absorbed locally (the OTS approximation).
The interior of the H~\textsc{ii} region is, however, optically thin to ionising radiation, and so this assumption breaks down towards the origin.
The good agreement with \textsc{Cloudy} in `Case B' mode lends support to this interpretation.
A more accurate radiative transfer scheme is beyond the scope of this work, but could potentially be achieved by coupling the multi-ion module to a radiative transfer scheme that treats every cell as a source, e.g., \textsc{TreeRay} \citep{WunWalDin21}.

There is a noticeable feature in the extension of a low level of He$^+$ ionisation beyond the ionisation front in \textsc{Pion} runs, and in \textsc{Cloudy} runs for test 3.
This is not related to charge exchange because the feature persists when this process is switched off.
Despite some effort we were unable to track down the source of this feature; we suspect it is related to radiative transfer of the hard photons and possibly sub-optimal arrangement of the energy bins.
Further investigation is deferred to a future work where we will implement a generalised OTS approximation including recombination radiation from, and ionisations of, all ions in the network.

\section{1D example application: Wind-wind interaction}\label{section:wind-wind}

We conducted a one-dimensional, spherically symmetric radiation-hydrodynamics (RHD) simulation to investigate the interaction between the stellar winds of a red supergiant (RSG) star that evolves to become a  Wolf-Rayet (WR) star when it loses its H-rich envelope. This simulation highlights the versatility of our module, effectively handling the distinctly different elemental abundances of the two winds and integrating all relevant physical processes, including photo-ionisation, collisional ionisation, recombination, and charge exchange reactions.

For this calculation we track 7 elements (H, He, C, N, O, Ne, Si) and their ionisation states by solving 51 ordinary differential equations and 7 constraint equations. The simulation used a spherically symmetric, nested grid with five levels, each containing 256 grid cells, with the coarsest level covering a radial domain of $3.7028\times10^{19}$ cm.
Each refined grid level, centered at the origin, has double the spatial resolution of the previous one and half of the domain size, with the same number of zones. The details of the 1D nested grid are described in \citet{Fichtner2024}.

The parameters defining both the stellar and circumstellar medium (CSM) are detailed in Table \ref{tab:wind-wind}. We model the CSM as being generated by the freely expanding stellar wind of a RSG with a terminal velocity, $\varv_{\rm \infty} = 25$\,km\,s$^{-1}$ and mass-loss rate $\dot{M}=2.0\times 10^{-5}$\,M$_{\odot}$\,yr$^{-1}$. The density is given by the conservation of mass:
\begin{equation}
\rho(r) = \frac{\dot{M}}{4\pi r^2 \varv_{\rm \infty}} \;.
\end{equation}
Thus, the CSM follows an inverse-square density distribution, and the elemental abundances are solar values from \citet{Asplund_2009}  (see Table \ref{tab:wind-wind}).

Further, the WR wind and CSM are photo-ionised by a central WR star of the carbon subclass (WC), characterized by a mass of \(11.6\) M\(_{\odot}\), an effective surface temperature of \(10^{5}\) K, a stellar radius of \(2.102\) R\(_{\odot}\), and a total luminosity of \(1.524 \times 10^{39}\)\,erg\,s$^{-1}$ (\(3.952\times10^{5}\,\mathrm{L}_{\odot}\)).
We adopt the Potsdam Wolf-Rayet model with Galactic WC abundances for the spectral energy distribution of the central star, which is divided into 15 energy bins in the energy range of 7.64 to 77.0 eV, as detailed in Table \ref{tab:bin-energy}.
The WR stellar wind is injected from the stellar surface with a mass-loss rate of $\dot{M}=3.0\times 10^{-5}$\,M$_{\odot}$\,yr$^{-1}$, and a wind terminal velocity of $\varv_{\rm \infty} = 1500$\,km\,s$^{-1}$ with 
chemical composition given in Table \ref{tab:wind-wind}.

\begin{table}
 \centering
  \caption{Parameters for 1D MHD Simulation of Wind-Wind Interaction: Ionising Source, Stellar Wind, and CSM.}
  \label{tab:wind-wind}
  \begin{tabular}{@{}l l@{}}
   \textbf{Stellar parameters (Unit)} & \textbf{Value} \\
   \hline
   WR $\textrm{T}_{\rm{eff}} \,\textrm{(K)}$ & $10^5$ \\
   WR ${R}_*\, (R_\odot)$   & 2.102 \\
   WR $L_* \, (L_{\odot})$& $3.952\times10^{5}$  \\
   WR SED & Potsdam Model WC\\
   $[\rm{M/H}]$ & 0.0 \\
   $\log g$ & 4.5\\
   WR $\dot{M}$  ($\mathrm{M}$$_\odot$ yr$^{-1}$) & $3.0\times 10^{-5}$ \\
   WR Wind $\varv_\infty$ (km/s) & 1500\\
   WR Wind X$_{\rm H}$ & $1.0\times 10^{-8}$ \\
   Wind X$_{\rm He}$ & $5.4541\times 10^{-1}$ \\
   Wind X$_{\rm C}$  & $4.0\times 10^{-1}$\\
   Wind X$_{\rm N}$  & $1.0\times 10^{-8}$\\
   Wind X$_{\rm O}$  & $5.0\times 10^{-2}$  \\
   Wind X$_{\rm Ne}$ & $1.88104\times 10^{-3}$ \\
   Wind X$_{\rm Si}$  & $7.72621\times 10^{-4}$ \\
     \hline
     \hline
   \textbf{CSM parameters (Unit)} & \textbf{Value} \\
 \hline
   RSG $\dot{M}$  ($\mathrm{M}$$_\odot$ yr$^{-1}$) & $2.0\times 10^{-5}$ \\
   RSG $\varv_\infty$ (km/s) & 25\\
   CSM X$_{\rm H}$  & $7.1263\times 10^{-1}$ \\
   CSM X$_{\rm He}$ & $2.7222\times 10^{-1}$ \\
   CSM X$_{\rm C}$  & $3.1049\times 10^{-3}$\\
   CSM X$_{\rm N}$  & $9.0990\times 10^{-4}$  \\
   CSM X$_{\rm O}$  & $8.4525\times 10^{-3}$  \\
   CSM X$_{\rm Ne}$ & $1.8977\times 10^{-3}$ \\
   CSM X$_{\rm Si}$ & $7.7906\times 10^{-4}$ \\
\hline
\end{tabular}
\end{table}

\begin{figure}
\includegraphics[width=\columnwidth]{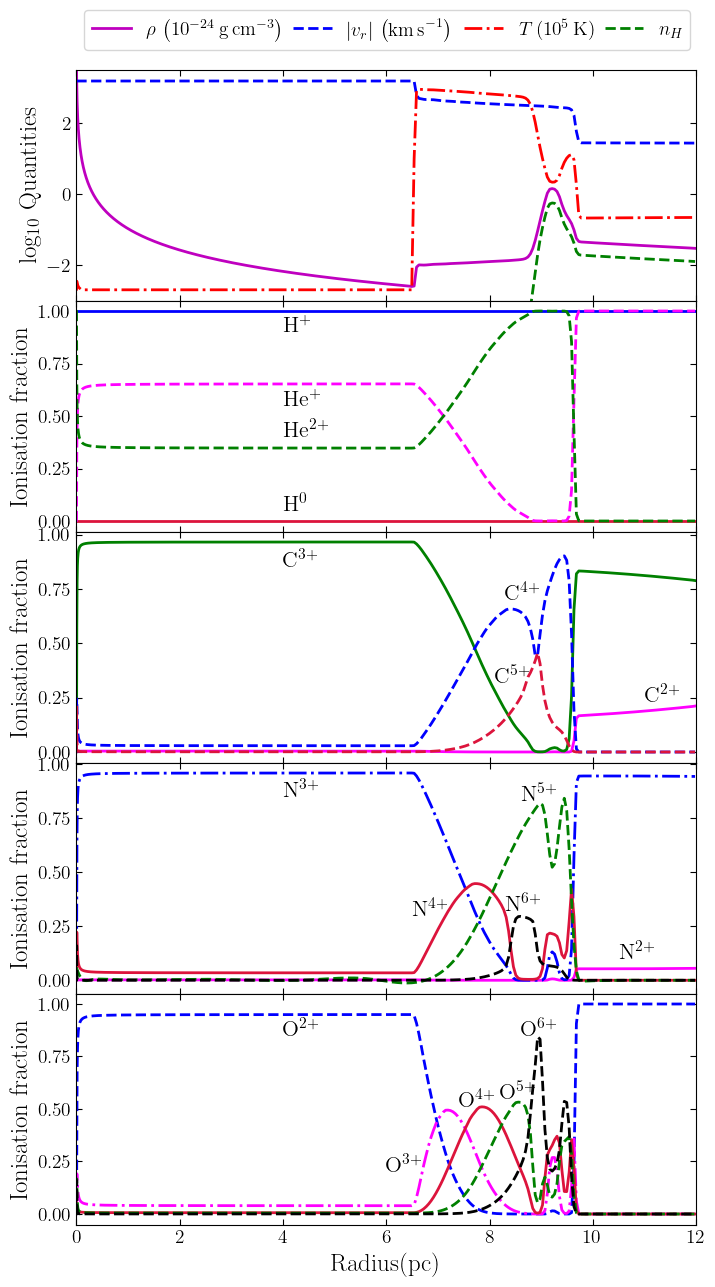}
\caption{
Snapshot from 1D simulation of wind-wind interaction from a WR wind expanding into a pre-existing RSG wind from a previous evolutionary phase.
The top panel displays the flow quantities, and the subsequent panels show the ionisation profiles of H, He, C, N, and O, all at a time 35\,kyr after the initiation of the WR wind.}
\label{fig:wind-wind-interaction}
\end{figure}

Figure \ref{fig:wind-wind-interaction} displays the flow quantities and ionisation profiles of H, He, C, N and O for the spherically symmetric interacting winds at \( t=3.5007 \times 10^4 \)\,yr.
The top panel shows the typical characteristics of a stellar wind bubble formed by the strong ram pressure of the central WR star wind, which sweeps up the slower wind from the RSG.
A key chemical feature distinguishing the WR wind from the RSG wind is the absence of H and N, demonstrated in the top panel by \( n_H \).
The rapidly increasing $n_\mathrm{H}$ at $r\approx9$\,pc marks the contact discontinuity between WR wind and RSG wind, mirrored by a rapid decrease in $T$ with $r$.
The finite width of this feature is a result of numerical diffusion in this relatively low-resolution simulation.

\subsection{Ionisation profile}

The stellar wind region is characterized by mechanical motion, with thermal energy remaining relatively low at around the minimum temperature in the simulation. ionisation in this area is entirely driven by photoionisation. Helium is almost entirely ionised, with over 63\% existing as \(\mathrm{He}^{2+}\) and the rest as \(\mathrm{He}^{1+}\). Carbon is primarily found as \(\mathrm{C}^{3+}\), making up about 95\% of all carbon in the stellar wind. Hydrogen is extremely scarce, with an abundance of only \(10^{-8}\) in the stellar winds of WC stars. Furthermore, the prevalent ionisation states of other elements include \(\mathrm{N}^{3+}\) and \(\mathrm{O}^{2+}\).

The expansion of the wind-wind interaction follows the constant velocity evolution described by \cite{KooMcK92}. The wind-wind interaction region is bounded by a forward shock expanding into the RSG wind and a reverse shock propagating into the WR wind. The reverse shock is adiabatic, whereas the forward shock is partially radiative. This region is of particular interest due to its large temperature range, spanning approximately \(2 \times 10^4\) K to \(8.8 \times 10^7\) K. The regions of shocked RSG wind and shocked WR wind are separated by a contact discontinuity, where the temperature sharply decreases from around \(2 \times 10^7\) K to \(2.7 \times 10^5\) K within a narrow region of width about 0.5 pc.

In the shocked WR wind region, temperatures range from roughly \(8.8 \times 10^7\) K to \(2 \times 10^7\) K over a span of about 2.3 pc. This area is dominated by highly ionised species, namely \(\mathrm{He}^{2+}\), \(\mathrm{C}^{4+}\), \(\mathrm{N}^{4+}\), \(\mathrm{N}^{5+}\), \(\mathrm{O}^{3+}\), \(\mathrm{O}^{4+}\), \(\mathrm{O}^{5+}\), and \(\mathrm{O}^{6+}\).
In this region, collisional ionisation overshadows photoionisation, indicating that the ionisation state of the gas is more significantly influenced by wind velocities and its chemical composition than by ionising photons from the central star. In the shock reference frame, protons moving at 1500\,km\,s$^{-1}$ have kinetic energy of nearly 12\,keV, much larger than the maximum photon energy considered for photoionisation (77\,eV), and so the ionisation state is determined by collisional processes.
Fig.~\ref{fig:wind-wind-interaction} shows that the ionisation timescale in this rarefied plasma is long enough that it is spatially resolved in the shocked WR wind.
We see successively higher-ionised states of C,N,O (and other elements not shown) up to the location of the contact discontinuity, with an ionisation length-scale of about a parsec, and the equilibrium ionisation states are not reached.

Conversely, the composition and temperature profile of the shell of the swept-up RSG wind are distinct from the shocked WR wind region. Although the temperature in this region remains high enough for collisional ionisation to dominate, the prevalent ions are different. The outer shock is partially radiative and expands at about 265\,km\,s$^{-1}$ into the RSG wind, at almost constant velocity according to \citet{KooMcK92}.
The relative velocity of the shock to the RSG wind is therefore 240\,km\,s$^{-1}$, and postshock temperature of $\sim0.3$\,keV.
The density in the shocked RSG wind is much larger than in the shocked WR wind, and so the ionisation timescale (and lengthscale) is much shorter.
This region exhibits a large fraction of \(\mathrm{H}^{1+}\) and \(\mathrm{He}^{2+}\), with \(\mathrm{C}^{4+}\), \(\mathrm{N}^{5+}\), and \(\mathrm{O}^{6+}\), especially around the contact discontinuity.

Furthermore, the ionisation of \(\mathrm{He}^{2+}\) drops immediately at the forward shock front. This phenomenon is attributed to the Strömgren sphere associated with \(\mathrm{He}^{2+}\) being located within the shocked wind region.

Beyond the forward shock the temperature is approximately \(20,000\) K, with the density of the RSG wind diminishing like \(r^{-2}\). A rough calculation shows that the Strömgren radius is on the order of kpc. This region is predominantly characterized by photoionisation, with the high temperature resulting from photo-heating. At a distance of 12 pc, the dominant ionised species are \(\mathrm{H}^{1+}\), \(\mathrm{He}^{1+}\), \(\mathrm{C}^{3+}\), \(\mathrm{N}^{3+}\), and \(\mathrm{O}^{2+}\). The RSG wind has been photoionised and photoheated to above the equilibrium temperature, but the cooling time is longer than the simulation runtime for this low-density outer-wind region.

\subsection{Line Luminosities}\label{section:wind-wind-linelumi}

\begin{figure}
\includegraphics[width=\columnwidth]{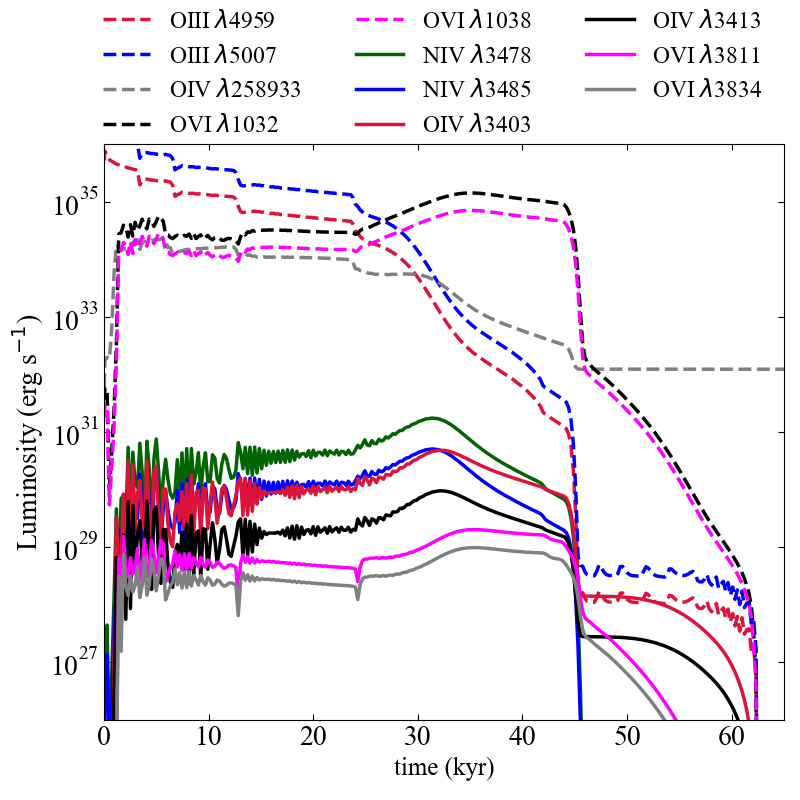}
\caption{Time evolution of luminosity for selected oxygen and nitrogen emission lines for the 1D wind-wind simulation in section~\ref{section:wind-wind}, calculated as descrbed in section~\ref{section:wind-wind-linelumi}.
The forward shock leaves the domain at around 45\,kyr, and the associated sharp drop in some line luminosities is therefore artificial.}
\label{fig:line-luminosity}
\end{figure}

The line radiation mechanism plays a key role in energy loss from hot plasma environments. The high-temperature regions resulting from wind-wind interactions provide an excellent context for analyzing the relationship between line emission, elemental abundances, and ionisation fractions.
Here, we calculated the time-dependent line luminosities for selected transitions of oxygen and nitrogen, as illustrated in Figure \ref{fig:line-luminosity}.
These luminosities were computed at each time step using the \textsc{CHIANTI} atomic database through the \textsc{CHIANTIPy} package v0.10.0 \citep{Zanna_2021, Young_2003}.
The wind-wind simulations provide ion density and temperature for each spherical cell, allowing us to compute the spectral line emissivity based on local cell temperature and to integrate it using the equation:
\begin{equation}
L_{\lambda} =  4\pi \int_V  \epsilon_{s,q}(\lambda, n_e, T)  \, n_{s, q} \, dV
\end{equation}
to derive the line luminosity at a specific time, where \(\epsilon_{s,q}(\lambda, n_e, T)\) represents the emissivity at wavelength \(\lambda\) and plasma electron density $n_e$ and temperature \(T\), with units of erg\,s$^{-1}$\,sr$^{-1}$\,ion$^{-1}_{s,q}$.

Fig.~\ref{fig:line-luminosity} shows the evolution of the luminosity of a range of strong nebular lines for the duration of the simulation.
We observe distinct behaviors among different spectral lines; however, lines with similar characteristics can be grouped together. For instance, the forbidden line doublet O{\sc iii}$\lambda\lambda$5007,4959, which are prominent nebular oxygen lines, exhibit similar trends. The luminosity of these lines continues to decrease as the nebula expands. This decline is primarily due to the decreasing mass density of the photoionised O{\sc iii} ion, which is prominent only in the unshocked WR and RSG wind regions. As the thin WR wind expands, displacing the denser RSG wind, the density of O {\sc iii} decreases accordingly. The abrupt drop in luminosity around \(t \sim 47\) kyr is attributed to the dense shocked shell leaving the simulation domain, leading to a sharp reduction in the total mass density.

The second group of lines, which exhibit similar trends, includes N{\sc iv}$\lambda\lambda$3478, 3485, O{\sc iv}$\lambda\lambda$3403, 3413, and O{\sc vi}$\lambda\lambda$3811, 3834. These lines are less luminous, with luminosities ranging from $10^{28}$ to $10^{31}$ erg s$^{-1}$. While N{\sc iv} is a dominant ion in the expanding WR and RSG wind regions, and O{\sc iv} and O{\sc vi} are confined to the shocked regions, they display comparable behavior. This similarity arises from the fact that N{\sc iv} also forms in the cooling shocked shell and has an ionisation fraction similar to that of O{\sc iv} and O{\sc vi} in this high-density region. This correlation suggests that the luminosity of these lines is primarily dictated by the density of the shocked shell.  The WR wind has a very low overall N abundance and so contributes little to the N{\sc iv} emissivity.

The most intriguing lines are O{\sc vi} $\lambda\lambda$1032, 1038, whose luminosity steadily increases up to 47 kyr and becomes the strongest of the lines that we calculate. This gradual rise is attributed to the increasing mass of the shocked shell as the WR wind sweeps up the RSG wind material. When the shocked shell exits the simulation domain, there is an abrupt decline in luminosity. Interestingly, after 47 kyr, the log luminosity follows a linear trend until 63 kyr, driven by the contribution of O{\sc vi} in the shocked WR wind region. This behavior indicates that the mass density associated with O{\sc vi} decreases exponentially over time within the simulation domain. Lastly, the O{\sc iv} $\lambda$258933 line, one of the prominent oxygen lines, exhibits a trend similar to that of O{\sc iii} $\lambda\lambda$5007, 4959. However, after around \(t \sim 47\) kyr, its luminosity stabilizes, remaining constant. This behavior is attributed to the unchanging density profile of O{\sc iv} ions in the freely expanding wind, which is sustained by photoionisation from the central star.

The line luminosity of some lines calculated and plotted in Fig.~\ref{fig:line-luminosity} show strong oscillations particularly at early times in the evolution of the nebula.
This arises because we do not spatially resolve the post-shock ionisation length of the forward shock at early times, and the emission from these lines is often dominated by only one or two cells in the post-shock region.
Depending on the position of the shock within a cell, the emission calculated for intermediate ionisation stages in the post-shock cell may vary substantially.
One can also see when the swept-up shell crosses grid-refinement boundaries at $t\approx 3, 6, 12, 24$\,kyr from small step changes in the luminosity of particularly O~\textsc{iii} lines.
These are related to adjustments in the hydrodynamic structure of the shocks and swept-up shell as they move to a region with larger numerical diffusivity.

\subsection{2D Wind-wind interaction}\label{section:2d-wind-wind}

We also perform two-dimensional radiation-hydrodynamics (RHD) simulations to model the interaction between the stellar winds of the RSG and WR stars.
These simulations are conducted in cylindrical coordinates, assuming rotational symmetry, using a computational grid in the $(R, z)$ plane.
The star is located at the origin, $(R, z) = (0, 0)$, within a rectangular simulation box.
The grid has a resolution of $128\times256$ cells at each level, and the coarsest level has a domain $R\in[0,3.7028\times10^{19}]$\,cm, and $z\in[-3.7028\times10^{19},3.7028\times10^{19}]$\,cm.
There are five grid levels, each centred at the origin, with each level having a spatial resolution twice as fine as the preceding coarse grid while maintaining the same number of zones, and with domain $2\times$ smaller in each dimension. 
The simulation parameters for both the stellar and circumstellar medium (CSM) are the same as those used in the one-dimensional simulations, as detailed in Table \ref{tab:wind-wind}, except that the wind boundary region (where the wind is injected) is set to 10 grid zones to reduce grid-related artefacts.
The simulation required about $10^4$ core-hours to complete, using 64 cores (8 MPI processes and 8 OpenMP threads per process).

\begin{figure}
\includegraphics[width=\columnwidth]{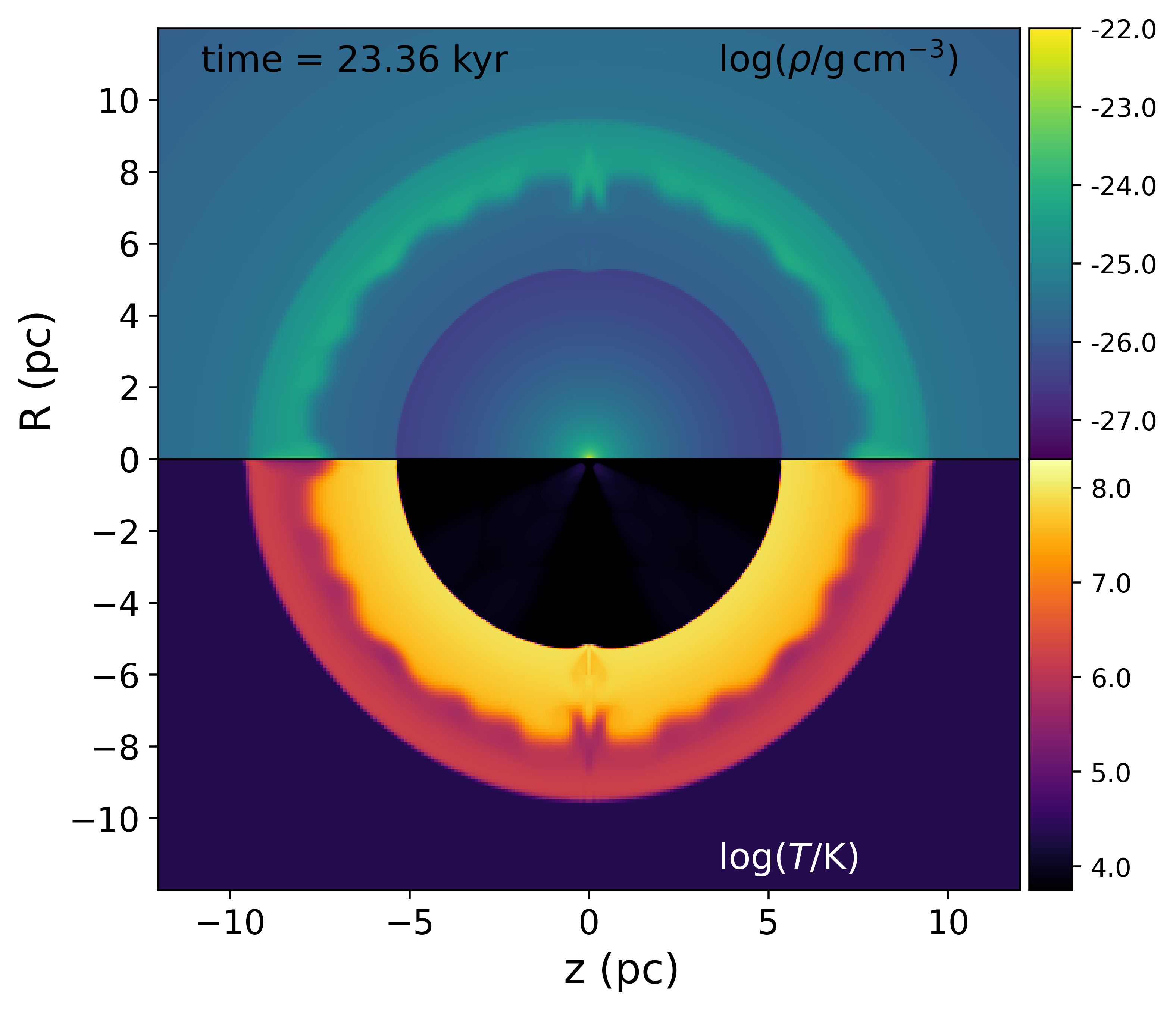}
\caption{Snapshot from a two-dimensional simulation captures the interaction between the WR wind and the remnant wind from the earlier RSG phase. The image shows the flow quantities 23.36 kyr after the onset of the WR wind, with the star located at the origin. The top panel shows $\log_{10}$ of gas density (g cm$^{-3}$) according to the colour scale shown, where brighter yellow regions indicate higher density, and darker purples represent lower density. The bottom panel shows $\log_{10}$ of temperature (in K) according to the colour scale displayed beside the plot.}
\label{fig:2dwind-flowquantities}
\end{figure}

\begin{figure*}
\includegraphics[width=\textwidth]{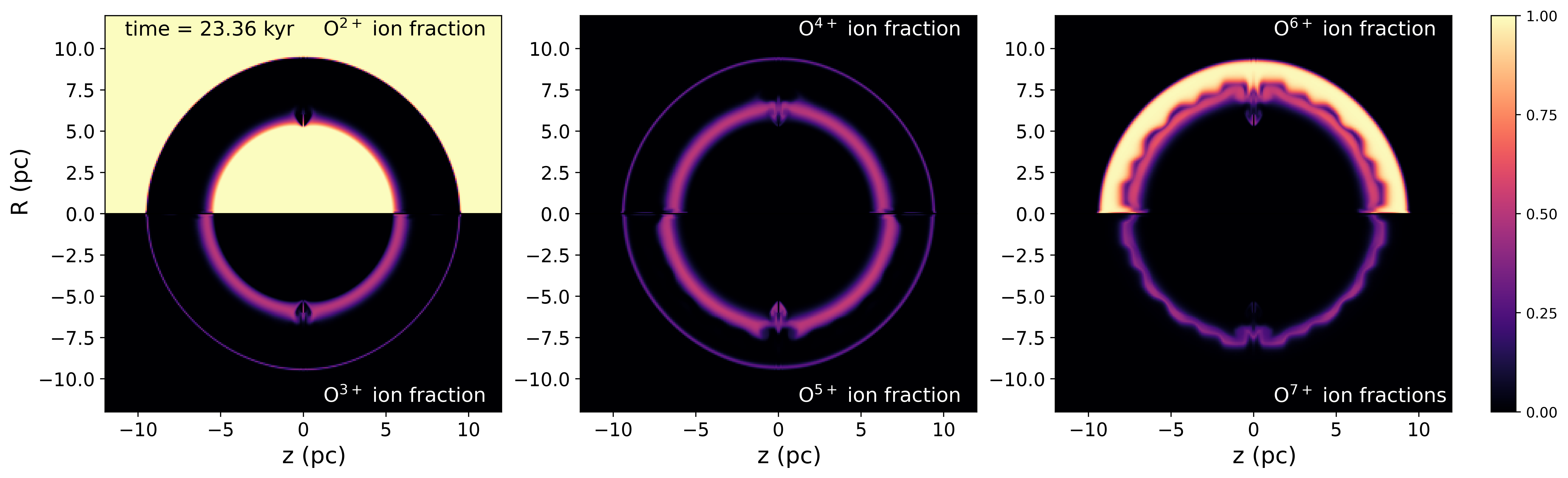}
Ionisation profile of O$^{2+}$ to O$^{7+}$ ions from a two-dimensional simulation of the interaction between the Wolf-Rayet (WR) wind and the remnant red supergiant (RSG) wind, shown 23.36 kyr after the onset of the WR wind.
The panels illustrate the ionisation fraction, $Y_{\mathrm{O}, i}$ on a linear scale, with white representing a oxygen completely in this ionisation state ($Y_{\mathrm{O}, i}$ = 1) and black denoting the zero abundance ($Y_{\mathrm{O}, i}$ = 0).
The colour scale is shown at the far right of the plot.
\label{fig:2dwind-ionfraction}
\end{figure*}

Fig.~\ref{fig:2dwind-flowquantities} shows a snapshot of the gas density and temperature on logarithmic scales, after about 23\,kyr of evolution.
The nebula evolves with almost spherical symmetry because the central star is able to fully ionise the whole domain very rapidly, so that D-type ionisation fronts (with their associated instabilities, \citealt{1996ApJ...469..171G}) do not develop.
Furthermore the wind termination shock is adiabatic, as seen in the 1D simulation, and the forward shock is also very close to adiabatic; there is a cooler, denser layer near the contact discontinuity but it is expanding faster than it can cool (also seen in 1D).
Some grid-related features are visible around the termination shock and contact discontinuity; these arise from the wind injection boundary and become less apparent with higher resolution.

Fig.~\ref{fig:2dwind-ionfraction} shows the ion fractions of the main species of oxygen in the domain (there is negligible O$^0$ or O$^{+}$ present because of the hard radiation field of the WR star).
The same features as from the 1D case (Fig.~\ref{fig:wind-wind-interaction}) are apparent: the freely expanding wind and the unshocked RSG wind are almost entirely in the form of O$^{2+}$, successively higher ionisation stages are reached at larger radius in the shocked WR wind, and the shocked RSG wind is primarily O$^{6+}$.
The shocked WR wind never reaches equilibrium ionisation (O$^{8+}$).
Overall the results agree very well with the 1D calculation.

\section{Discussion}\label{section:discussion}

\subsection{Simulation Performance}\label{subsection:perfo}

\begin{figure}
\includegraphics[width=\columnwidth]{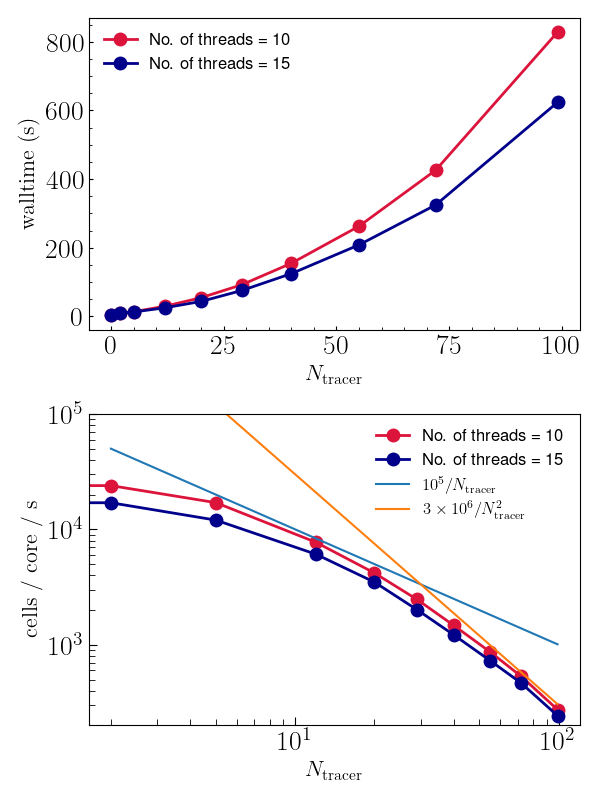}
\caption{Relation between simulation speed and the number of tracers (elements plus their ions). The upper plot shows the walltime of a 1D simulation using from 1 to 9 elements, with either 10 or 15 OpenMP threads on 10 or 15 cores.  The lower panel shows the same data, calculating the number of of cells updated per core per second.  This 1D shock calculation did not include photoionisation or charge-exchange reactions (see text for details).}
\label{fig:walltime_cellupdatespeed}
\end{figure}

\begin{figure}
\includegraphics[width=\columnwidth]{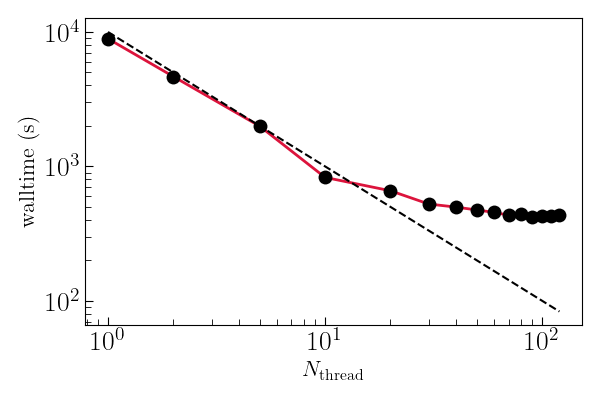}
\caption{Parallel scaling of the same 1D shock test as Fig.~\ref{fig:walltime_cellupdatespeed} (except that now charge-exchange reactions are included), showing the simulation walltime as a function of the number of OpenMP threads (and cores) used for the calculation.
This is compared with the ideal speedup of $1/N_\mathrm{thread}$ (dashed line).}
\label{fig:walltime_nthreads}
\end{figure}

Here we investigate the impact of chemical tracers on the computational efficiency of MHD simulations.
We use a modified version of the Raymond planar shock model E, changed to have an inflow rate of 1000 kms$^{-1}$ and a transverse magnetic field of $10^{-6}$ G.
We perform several simulations with sequential addition of tracers, including H, He, C, N, O, Ne, Si, S, and Fe, along with their respective ionisation states for a fixed simulation time of $3.156\times10^9$ s, using a uniform 1D grid of 256 grid zones and slab symmetry.

Figure \ref{fig:walltime_cellupdatespeed} shows the walltime and cell update-speed as a function of the function of number of tracers, $N_\mathrm{tracer}$, for runs with 10 and 15 OpenMP threads.
For this test we included only collisional processes excluding charge exchange. The walltime required is the sum of the microphysics update, the MHD update, and the boundary updates, but the runtime is quickly dominated by the microphysics solver as $N_\mathrm{tracer}$ increases.
For large $N_\mathrm{tracer}$ it looks like the update speed approaches a $N_\mathrm{tracer}^{-2}$ scaling.

We also examine how the runtime scales with the number of OpenMP threads, $N_\mathrm{thread}$, in the same shock test, but this time including the charge-exchange reactions with \( q = 4 \). The results are shown in Figure \ref{fig:walltime_nthreads}.

The walltime decreases as $N_\mathrm{thread}^{-1}$ up to 10 OpenMP threads and there is little gain beyond going to 30 threads.
This is partly because for 1D tests the MHD and boundary updates are not multi-threaded. Each OpenMP thread operates on a 1D column of cells for MHD, whereas the threads operate on a cell-by-cell basis for the microphysics update.

Overall the code performs as expected and, when running with all 9 elements and 99 tracers, the update speed is about $1000\times$ slower than an MHD calculation without microphysics.
This means that, while 2D and 3D simulations are still feasible with the module, they must be significantly lower resolution than corresponding simulations without non-equilibrium ionisation.
Three possible avenues to speed up the calculation, which will be explored in future work, are:

\begin{enumerate}
    \item 
    Update cells in batches, taking advantage of vectorization that is possible with \textsc{Sundials}.
    This requires changes to the module data structures for both the state vectors and the extra parameters such as incident radiation field.
    \item 
    Currently we solve using an approximate Jacobian calculated internally by \textsc{Sundials}, but for problems of this size we should see significant performance gains by implementing a function to calculation an analytic Jacobian matrix.
    This is not trivial because of the radiation field and the possibility that cells are optically thick (so the sensitivity of ionisation rates to other species depends on optical depth), but it should improve performance.
    \item 
    Offload some calculation on GPUs, likely in combination with batch integration of cells.
    This has recently been demonstrated to provide efficiency gains by \citet{BalDayEsc24} and should be feasible for this module.
\end{enumerate}

\subsection{Importance of non-equilibrium ionisation}

\begin{table}
\centering
\caption{Comparison of line luminosities for selected oxygen and nitrogen lines under non-equilibrium and equilibrium conditions. The second and third columns show the luminosity values for non-equilibrium and equilibrium states, respectively. The fourth column indicates the ratio of line luminosity in the non-equilibrium state relative to the equilibrium state.}
\renewcommand{\arraystretch}{1.1} 
\label{tab:luminosity_neq_ieq}
\begin{tabular}{lccc}
\textbf{Line} & \multicolumn{2}{c}{\textbf{L$_\lambda$} ($10^{32}$ \rm{erg \, s}$^{-1}$)} & \textbf{L$_\lambda^{NEQ}$/L$_\lambda^{IEQ}$} \\ 
\hline
& NEQ  & IEQ & \\  
\hline
O\, {\sc iii} $\lambda$4959 &2.7285E+0&3.5581E+0& 0.7668\\	 
O\, {\sc iii} $\lambda$5007	&7.8884E+0&1.0287E+1& 0.7668\\

O\, {\sc iv} $\lambda$258933	&1.4367E+1&2.0690E+1& 0.6944\\ 
O\, {\sc iv} $\lambda$3403 &3.0873E$-2$&4.3716E$-2$& 0.7062\\ 
O\, {\sc iv} $\lambda$3413	&6.1029E$-3$&8.6416E$-3$& 0.7062\\ 

O\, {\sc vi} $\lambda$1032	&1.2330E+3&8.9595E+2& 1.3762\\
O\, {\sc vi} $\lambda$1038  &6.1340E+2&4.4536E+2& 1.3773\\
O\, {\sc vi} $\lambda$3811&1.6685E$-3$&6.4220E$-4$& 2.5981\\
O\, {\sc vi} $\lambda$3834&8.1111E$-4$&3.1059E$-4$& 2.6115\\

O\, {\sc vii} $\lambda$21.602&5.0408E+0&8.3881E$-1$& 6.0095\\
O\, {\sc vii} $\lambda$21.804&4.6634E$-1$&1.6723E$-1$& 2.7884\\
O\, {\sc vii} $\lambda$22.098&4.5450E+0&7.7364E$-1$& 5.8748\\

O\, {\sc viii} $\lambda$18.973&1.2915E$-1$&1.3812E$-1$& 0.9351\\
O\, {\sc viii} $\lambda$18.967&2.5843E$-1$&2.7631E$-1$& 0.9353\\

N\, {\sc iv} $\lambda$3478 &6.0819E$-2$&5.0773E$-2$& 1.1979\\	
N\, {\sc iv} $\lambda$3485&1.7605E$-2$&1.4698E$-2$& 1.1978\\	
\hline
\end{tabular}
\end{table}

\begin{figure*}[ht]
\centering
\includegraphics[width=\textwidth]{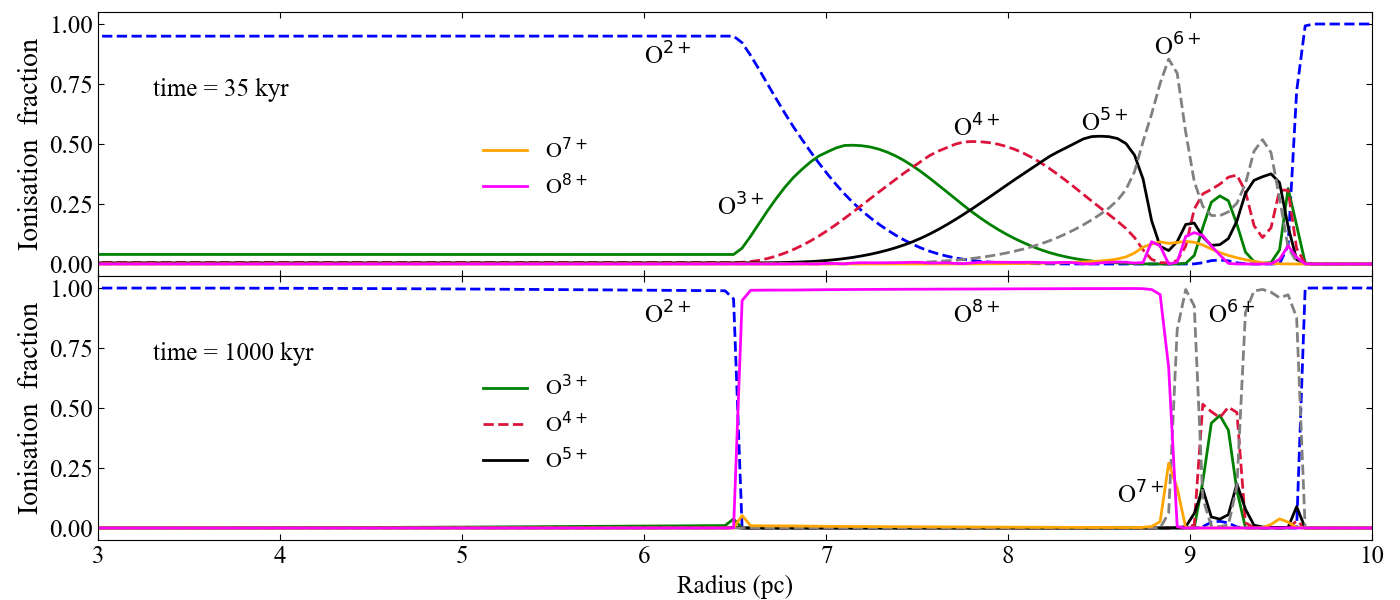}
\caption{Ionisation profiles of oxygen. The upper panel illustrates the NEQ case at \(t = 35\) kyr, while the lower panel shows the IEQ state at \(t = 1000\) kyr.}
\label{figure:NEQ-CIE}
\end{figure*}

In this subsection, we explore the significance of non-equilibrium ionisation (NEQ) by conducting a comparative analysis between the ionisation profiles obtained from wind-wind interactions of the RSG and WR simulations (described in Section \ref{section:wind-wind}) at $t=35$ kyr, representing a non-equilibrium state, and those derived under equilibrium conditions (ionisation equilibrium is denoted IEQ).
The IEQ state is achieved by evolving the simulation from $t=35$ kyr to $1000$ kyr, with hydrodynamics turned off and the energy equation $\dot{E}$ set to zero in the microphysics module.
After 1000 kyr, the ionisation profiles of wind-wind interactions between the RSG and WR have relaxed to a steady state in all regions of the flow. Figure~\ref{figure:NEQ-CIE} presents a comparison between the NEQ ionisation profile of oxygen in the top panel and its IEQ counterpart in the bottom panel. 

There is a slight but noticeable difference in the ionisation of the gas in the freely expanding WR stellar wind region ($r<6.5$\,pc) beteween the NEQ and IEQ profiles, in that the NEQ calculation has about 5 per cent \(\mathrm{O}^{3+}\) and 95 per cent \(\mathrm{O}^{2+}\), whereas the IEQ calculation is effectively 100 per cent \(\mathrm{O}^{2+}\). This difference arises because ionisation in this region is largely driven by photoionisation, and the photoionisation equilibrium timescale is longer than the expansion timescale of the wind.

In the shocked WR wind region (approximately $r\in[6.5,8.8]$\,pc, cf.~Fig.~\ref{fig:wind-wind-interaction}) where temperatures range from approximately \(2 \times 10^7\)~K to \(8.8 \times 10^7\)~K, oxygen is entirely in the \(\mathrm{O}^{8+}\) state in the equilibrium case, a state that is less prevalent in the NEQ scenario.
Since this region is dominated by collisional ionisation (thermal energy per particle is much larger than the photon energy from the stellar SED), the IEQ profile aligns with the one obtained under CIE conditions. For this temperature range, the CIE results show fully ionised oxygen, consistent with the IEQ profile obtained here. The NEQ profile can be understood based on the post-shock velocity ($v_\mathrm{ps}\approx1500/4$\,km\,s$^{-1}$), temperature ($T\approx 10^8$\,K) and electron density ($n_e\sim0.003\,\mathrm{cm}^{-3}$).
For this temperature the collisional ionisation rate-coefficient for O$^{2+}$ is $\zeta_{\rm{O},2}^\mathrm{coll}\approx10^{-8}$\,cm$^{3}$\,s$^{-1}$, giving an ionisation length scale $\ell_\mathrm{i}= v_\mathrm{ps} / (n_e \zeta_{\rm{O},2}^\mathrm{coll}) \approx 0.3$\,pc.
Fig.~\ref{figure:NEQ-CIE} shows that the length scale for the O$^{2+}$ abundance to decrease to half of its pre-shock value is $\approx0.25$\,pc, comparable to our rough estimate.
At this temperature, $\zeta_{\rm{O},2}^\mathrm{coll}\approx2\zeta_{\rm{O},3}^\mathrm{coll}$, so the $\rm{O}^{3+}\rightarrow \rm{O}^{4+}$ transition has a somewhat larger length scale (the slower rate is somewhat compensated by the flow velocity decreasing with increasing $r$).
In the NEQ calculation, the post-shock WR wind never reaches full ionisation over a length scale of nearly 2.5\,pc.
This is likely not directly observable because the density (and emission measure) of the gas is so low, but it may have an observable effect at the contact discontinuity where thermal conduction and turbulent mixing produce denser gas that can be detected in soft X-rays \citep{ToaArt18}.

The IEQ profile of the swept-up RSG wind differs from the corresponding NEQ case, although the dominant ion remains the same.
In the NEQ scenario, the region is primarily composed of \(\mathrm{O}^{6+}\) and \(\mathrm{O}^{5+}\), with \(\mathrm{O}^{6+}\) slightly prevailing over \(\mathrm{O}^{5+}\).
In contrast, the equilibrium case is dominated solely by \(\mathrm{O}^{6+}\), with some O$^{4+}$ and O$^{3+}$ in the cooling layer behind the shock. This observation aligns with the CIE ionisation results (see Figure~\ref{figure:CIE}), where the \(\mathrm{O}^{6+}\) ion fraction dominates within the temperature range of \(3 \times 10^5\)~K to \(2 \times 10^6\)~K.
In the NEQ case the relatively long ionisation timescale results in significant O$^{4+}$ and O$^{5+}$ in the postshock layer, together with O$^{6+}$ with almost equal abundances.

We further compare the line luminosities between the NEQ and IEQ cases. Given that the density and temperature profiles remain identical in both scenarios, the ion fraction determines the line luminosity.
As discussed in Section \ref{section:wind-wind-linelumi}, the luminosity is predominantly determined by the mass density of the specific ion within the shell volume.
Therefore, we anticipate distinct luminosities, as the ionisation profiles differ significantly in this region for both cases.
Table \ref{tab:luminosity_neq_ieq} presents a comparison of line luminosities for the same transitions discussed in Section \ref{section:wind-wind-linelumi} for both the NEQ and IEQ cases.
The O\,{\sc iii} and O\,{\sc iv} lines exhibit lower luminosities in NEQ compared to IEQ, whereas the O\,{\sc vi} and N\,{\sc iv} lines show higher luminosities in NEQ.
Notably, the O\,{\sc vi} $\lambda$3811 and O\,{\sc vi} $\lambda$3834 transitions display luminosities that are approximately 2.5 times greater in NEQ.
The reduced luminosities of O\,{\sc iii} and O\,{\sc iv} in NEQ are due to their higher mass densities in IEQ, as their ion fractions peak in regions of highest gas density.
Conversely, the enhanced luminosities of O\,{\sc vi} and N\,{\sc iv} in NEQ are explained by their increased mass densities under non-equilibrium conditions.

\subsection{Electron-ion temperature equilibration}

We should caution that in the X-ray emitting regime, our single-fluid assumption that the electrons and ions have the same temperature may not be correct in the post-shock plasma.
In collisionless shocks ions are heated more than electrons \citep{Spi62}, although the degree of the difference is still not well understood \citep{GhaSchMit13} and measurements from supernova remnant shocks give different results compared with Solar Wind measurements and PIC simulations \citep{RayGhaBoh23}.
Nevertheless, the temperature ratio $T_e/T_i$ can be as low as 0.05$-$0.1 in supernova remnant shocks \citep{RayGhaBoh23}.
The electron-ion temperature equilibration via Coulomb collisions with protons has a timescale \citep{Spi62,GhaSchMit13}
\begin{equation}
    \tau_{ei} \approx 240\,\mathrm{yr} \left(\frac{v_\mathrm{sh}}{1500\,\mathrm{km\,s}^{-1}}\right)^3 \left(\frac{0.01 \mathrm{cm}^{-3}}{n_e}\right) \;,
\end{equation}
on substituting $\ln \Lambda=30$ for the Coulomb logarithm.
For heavier ions $\tau_{ei}$ is multiplied by $m_i/m_p Z_i^2$ \citep[e.g.][]{PolCorSte05}, where $m_i$ and $Z_i$ are the mass and charge of ion $i$, respectively (i.e., the timescale is shorter for heavy ions).
This gives a length-scale in the shocked WR wind of the wind-wind simulation of $\ell_{ei}\approx0.3$\,pc, comparable to the ionisation length-scale calculated above.
Thus we may expect that X-ray emission from post-shock gas from fast shocks may not be accurately modelled in low-density plasma in the single-temperature approximation, and similarly the line emission in the soft X-ray O~\textsc{vii} and O~\textsc{viii} lines.
Observations show that for shock speeds $v_\mathrm{sh}\lesssim400$\,km\,s$^{-1}$ the electrons and ions have the same post-shock temperature \citep{GhaSchMit13}.

\section{Conclusions}\label{section:conclusion}

After extensive testing and validation, we have successfully integrated a solver for multi-ion non-equilibrium ionised plasmas into the radiation-MHD code, \textsc{pion}.
This enhancement equips \textsc{pion} with the capability to spatially and temporally determine the distribution of ionisation levels for key elements -- H, He, C, N, O, Ne, Si, S, and Fe -- essential for radiative cooling and spectral line diagnostics.
The solver incorporates collisional ionisation, photoionisation, charge-exchange reactions with H and He, and cosmic-ray ionisation of neutral species.
It supports arbitrary elemental abundances for neutral and ionised gas (excluding molecular forms) and includes an enhanced non-equilibrium radiative cooling calculation.
The ion-by-ion cooling curve, calculated based on instantaneous ion fractions, will facilitate more accurate modeling of ionisation in astrophysical scenarios where non-equilibrium ionisation conditions occur.
This capability allows the module to predict spectral lines in the UV, optical, IR, and X-ray bands, even when the plasma is out of ionisation equilibrium, thus enabling more realistic comparison of simulations with observational data.

Various tests have been conducted to verify the solver's accuracy.
Collisional ionisation equilibrium was achieved, with the solver converging to the correct ionisation state, consistent with the results previously reported by, e.g., \citet{Sutherland_1993}. 
The solver reproduced the CIE cooling curve for both solar and Wolf-Rayet (WC) abundances, matching the findings reported by \citet{Sutherland_1993} and \citet{Eatson_2022}, respectively.
When the same tests were performed with charge exchange enabled, the ionisation profiles displayed notable differences, closely matching the findings in the works of \cite{Gu2022} and \cite{ArnaudRothenflug1985}. 
Further validation was done by comparing the module's results with both equilibrium and non-equilibrium calculations from existing literature.
Several shock models, akin to Model E in \citep{Raymond1979}, were simulated, including both adiabatic and non-adiabatic scenarios.
Despite considerable resolution differences, the adiabatic shock tests showed excellent consistency in the ionisation profiles across low and high-resolution simulations.
Non-adiabatic shock tests highlighted the non-equilibrium ionisation structure of the post-shock gas during radiative cooling and demonstrated the impact of charge exchange on ion abundances in cooling plasmas.
Additionally, simulations of photoionised H~\textsc{ii} regions were conducted to validate the photoionisation subroutine in a multi-ion environment, following the HII40 Lexington benchmark and showing good overall agreement with results from the photoionisation code \textsc{Cloudy}.

We conducted a comprehensive analysis of the performance and scalability of the MHD code with the new microphysics solver.
The solver requires more computation with increasing \(N_\mathrm{tracer}\), with the runtime becoming predominantly influenced by the microphysics solver for \(N_\mathrm{tracer}\gg10\) and the cell update speed approaches a \(N_\mathrm{tracer}^{-2}\) scaling.
The walltime decreases significantly with increasing OpenMP thread count, adhering to a \(N_\mathrm{thread}^{-1}\) pattern up to 10$-$20 threads, and continues to show efficiency gains even with up to 30 threads.
Overall, the code performance is commendable, facilitating detailed 2D simulations and enabling low-resolution 3D simulations.
There is room for performance improvements by multithreading the microphysics integrator (\textsc{cvode}) with OpenMP or GPU acceleration, and this will be explored in future work.
The multi-ion module will form part of the next release of \textsc{pion} in 2025.

To demonstrate the capabilities of \texttt{NEMO}, we conducted 1D and 2D radiation hydrodynamics (RHD) simulations of the time-dependent spherical expansion of a Wolf-Rayet wind into a pre-existing red supergiant wind, with strongly differing elemental abundances in the two winds, including photo-ionisation, collisional ionisation, recombination, and charge exchange reactions.
We investigated the difference between our non-equilibrium results and the ionisation structure when integrated to equilibrium.
There are significant differences that also give different line luminosities for bright nebular emission lines from soft X-ray to thermal IR.
This application underscores the versatility of our module and showcases its effectiveness in modeling various ionisation processes in distinct regions.
In future work we will apply this module to study supernova remnants, nebulae around Wolf-Rayet stars and shocked plasma in colliding-wind binaries.

\begin{acknowledgements}
AM expresses his gratitude to Dr. Palmira Jim\'enez-Hern\'andez
 for guiding him in the use of \textsc{Cloudy}, which enabled him to obtain ionisation profiles of HII regions.
AM acknowledges support from a Royal Society Enhanced Research Expenses grant (RF\textbackslash ERE\textbackslash 210382).
JM acknowledges support from a Royal Society--Science Foundation Ireland University Research Fellowship.
MC acknowledges support from the Gates Cambridge Scholarship.
TJH acknowledges funding from a Royal Society Dorothy Hodgkin Fellowship and UKRI guaranteed funding for a Horizon Europe ERC consolidator grant (EP/Y024710/1).
\end{acknowledgements}

\bibliographystyle{aa}
\bibliography{references} 

\end{document}